\documentclass[trackchanges]{aastex701}
\usepackage[utf8]{inputenc}  
\usepackage[T1]{fontenc}     

\usepackage{amsmath}
\usepackage{amssymb}
\usepackage{amsfonts}
\usepackage{amsthm}
\usepackage{mathtools}
\usepackage{bm}
\usepackage{upgreek}
\usepackage{mathrsfs}

\usepackage{graphicx}
\usepackage{float}
\usepackage{caption}
\usepackage{subcaption}  
\usepackage{overpic}

\usepackage{array}
\usepackage{tabularx}
\usepackage{longtable}
\usepackage{booktabs}
\usepackage{multirow}
\usepackage{makecell}
\usepackage[flushleft]{threeparttable}  

\usepackage{xcolor}
\usepackage{hyperref}  
\usepackage[all]{hypcap}
\usepackage{url}

\usepackage{textcomp}
\usepackage{pifont}
\usepackage{xspace}
\usepackage{soul}
\usepackage{enumitem}
\usepackage{csquotes}
\usepackage{rotating}
\usepackage{mwe,tikz}
\usepackage{epstopdf}  

\usepackage{algorithm}
\usepackage{algorithmicx}
\usepackage{algpseudocode}

\usepackage[title]{appendix}

\usepackage{subfiles}

\graphicspath{{./}}

\begin{document}

\title{Illuminating the Mass Gap Through Deep Optical Constraint on a Neutron Star Merger Candidate S250206dm}

\correspondingauthor{Wen Zhao, Zhiping Jin, Zigao Dai}
\email{wzhao7@ustc.edu.cn, jin@pmo.ac.cn, daizg@ustc.edu.cn}

\author[0000-0002-2242-1514,gname=Zhengyan,sname=Liu]{Zhengyan Liu}
\altaffiliation{These authors contributed equally to this work.}
\affiliation{Department of Astronomy, University of Science and Technology of China, Hefei 230026, China}
\affiliation{School of Astronomy and Space Sciences, University of Science and Technology of China, Hefei 230026, China}
\email{}

\author{Zelin Xu}
\altaffiliation{These authors contributed equally to this work.}
\affiliation{Department of Astronomy, University of Science and Technology of China, Hefei 230026, China}
\affiliation{School of Astronomy and Space Sciences, University of Science and Technology of China, Hefei 230026, China}
\email{}

\author[0000-0002-9092-0593]{Ji-an Jiang}
\affiliation{Department of Astronomy, University of Science and Technology of China, Hefei 230026, China}
\affiliation{School of Astronomy and Space Sciences, University of Science and Technology of China, Hefei 230026, China}
\affiliation{National Astronomical Observatory of Japan, National Institutes of Natural Sciences, Tokyo 181-8588, Japan}
\email{}

\author[0000-0002-1330-2329]{Wen Zhao}
\affiliation{Department of Astronomy, University of Science and Technology of China, Hefei 230026, China}
\affiliation{School of Astronomy and Space Sciences, University of Science and Technology of China, Hefei 230026, China}
\email{wzhao7@ustc.edu.cn}

\author[0000-0003-4977-9724]{Zhiping Jin}
\affiliation{Purple Mountain Observatory, Chinese Academy of Sciences, Nanjing 210023, China}
\email{jin@pmo.ac.cn}

\author[0000-0002-7835-8585]{Zigao Dai}
\affiliation{Department of Astronomy, University of Science and Technology of China, Hefei 230026, China}
\affiliation{School of Astronomy and Space Sciences, University of Science and Technology of China, Hefei 230026, China}
\email{daizg@ustc.edu.cn}

\author{Dezheng Meng}
\affiliation{Department of Astronomy, University of Science and Technology of China, Hefei 230026, China}
\affiliation{School of Astronomy and Space Sciences, University of Science and Technology of China, Hefei 230026, China}
\email{}

\author[0000-0002-6299-1263]{Xuefeng Wu}
\affiliation{Purple Mountain Observatory, Chinese Academy of Sciences, Nanjing 210023, China}
\email{}

\author[0000-0002-9758-5476]{Daming Wei}
\affiliation{Purple Mountain Observatory, Chinese Academy of Sciences, Nanjing 210023, China}
\email{}

\author[0000-0001-6223-840X]{Runduo Liang}
\affiliation{Department of Astronomy, University of Science and Technology of China, Hefei 230026, China}
\affiliation{School of Astronomy and Space Sciences, University of Science and Technology of China, Hefei 230026, China}
\affiliation{National Astronomical Observatories, Chinese Academy of Sciences, Beijing 100101, China}
\email{}

\author[0000-0001-7613-5815]{Lei He}
\affiliation{Department of Astronomy, University of Science and Technology of China, Hefei 230026, China}
\affiliation{School of Astronomy and Space Sciences, University of Science and Technology of China, Hefei 230026, China}
\email{}

\author[0000-0003-4721-6477]{Minxuan Cai}
\affiliation{Department of Astronomy, University of Science and Technology of China, Hefei 230026, China}
\affiliation{School of Astronomy and Space Sciences, University of Science and Technology of China, Hefei 230026, China}
\email{}

\author[0000-0003-4200-4432]{Lulu Fan}
\affiliation{Department of Astronomy, University of Science and Technology of China, Hefei 230026, China}
\affiliation{School of Astronomy and Space Sciences, University of Science and Technology of China, Hefei 230026, China}
\affiliation{Institute of Deep Space Sciences, Deep Space Exploration Laboratory, Hefei 230026, China}
\email{}

\author{Weiyu Wu}
\affiliation{Department of Astronomy, University of Science and Technology of China, Hefei 230026, China}
\affiliation{School of Astronomy and Space Sciences, University of Science and Technology of China, Hefei 230026, China}
\email{}

\author{Junhan Zhao}
\affiliation{Department of Astronomy, University of Science and Technology of China, Hefei 230026, China}
\affiliation{School of Astronomy and Space Sciences, University of Science and Technology of China, Hefei 230026, China}
\email{}

\author{Ziqing Jia}
\affiliation{Department of Astronomy, University of Science and Technology of China, Hefei 230026, China}
\affiliation{School of Astronomy and Space Sciences, University of Science and Technology of China, Hefei 230026, China}
\email{}

\author{Kexin Yu}
\affiliation{Department of Astronomy, University of Science and Technology of China, Hefei 230026, China}
\affiliation{School of Astronomy and Space Sciences, University of Science and Technology of China, Hefei 230026, China}
\email{}

\author[0000-0001-9648-7295]{Jinjun Geng}
\affiliation{Purple Mountain Observatory, Chinese Academy of Sciences, Nanjing 210023, China}
\email{}

\author{Di Xiao}
\affiliation{Purple Mountain Observatory, Chinese Academy of Sciences, Nanjing 210023, China}
\email{}

\author{Feng Li}
\affiliation{State Key Laboratory of Particle Detection and Electronics, University of Science and Technology of China, Hefei 230026, China}
\email{}

\author{Jinlong Tang}
\affiliation{Institute of Optics and Electronics, Chinese Academy of Sciences, Chengdu 610209, China}
\email{}

\author{Yingxi Zuo}
\affiliation{Purple Mountain Observatory, Chinese Academy of Sciences, Nanjing 210023, China}
\email{}

\author{Xiaoling Zhang}
\affiliation{Purple Mountain Observatory, Chinese Academy of Sciences, Nanjing 210023, China}
\email{}

\author{Hao Liu}
\affiliation{State Key Laboratory of Particle Detection and Electronics, University of Science and Technology of China, Hefei 230026, China}
\email{}

\author[0000-0003-1617-2002]{Jian Wang}
\affiliation{Institute of Deep Space Sciences, Deep Space Exploration Laboratory, Hefei 230026, China}
\affiliation{State Key Laboratory of Particle Detection and Electronics, University of Science and Technology of China, Hefei 230026, China}
\email{}

\author[0000-0002-1463-9070]{Hongfei Zhang}
\affiliation{State Key Laboratory of Particle Detection and Electronics, University of Science and Technology of China, Hefei 230026, China}
\email{}

\author{Ming Liang}
\affiliation{National Optical Astronomy Observatory (NSF's National Optical-Infrared Astronomy Research Laboratory), 950 N Cherry Ave, Tucson Arizona 85726, USA}
\email{}

\author[0000-0002-4372-0759]{Hairen Wang}
\affiliation{Purple Mountain Observatory, Chinese Academy of Sciences, Nanjing 210023, China}
\email{}

\author{Dazhi Yao}
\affiliation{Purple Mountain Observatory, Chinese Academy of Sciences, Nanjing 210023, China}
\email{}

\author[0000-0001-7201-1938]{Lei Hu}
\affiliation{McWilliams Center for Cosmology, Department of Physics, Carnegie Mellon University, 5000 Forbes Ave, Pittsburgh, 15213, PA, USA}
\affiliation{Purple Mountain Observatory, Chinese Academy of Sciences, Nanjing 210023, China}
\email{}

\author[0000-0002-7660-2273]{Xu Kong}
\affiliation{Department of Astronomy, University of Science and Technology of China, Hefei 230026, China}
\affiliation{School of Astronomy and Space Sciences, University of Science and Technology of China, Hefei 230026, China}
\affiliation{Institute of Deep Space Sciences, Deep Space Exploration Laboratory, Hefei 230026, China}
\email{}

\author[0000-0001-9327-0920]{Bin Li}
\affiliation{Purple Mountain Observatory, Chinese Academy of Sciences, Nanjing 210023, China}
\email{}

\author[0000-0002-7152-3621]{Ning Jiang}
\affiliation{Department of Astronomy, University of Science and Technology of China, Hefei 230026, China}
\affiliation{School of Astronomy and Space Sciences, University of Science and Technology of China, Hefei 230026, China}
\email{}

\author[0000-0002-1517-6792]{Tinggui Wang}
\affiliation{Department of Astronomy, University of Science and Technology of China, Hefei 230026, China}
\affiliation{School of Astronomy and Space Sciences, University of Science and Technology of China, Hefei 230026, China}
\affiliation{Institute of Deep Space Sciences, Deep Space Exploration Laboratory, Hefei 230026, China}
\email{}

\author[0000-0002-3105-3821]{Zhen Wan}
\affiliation{Department of Astronomy, University of Science and Technology of China, Hefei 230026, China}
\affiliation{School of Astronomy and Space Sciences, University of Science and Technology of China, Hefei 230026, China}
\email{}

\author[0000-0002-1935-8104]{Yongquan Xue}
\affiliation{Department of Astronomy, University of Science and Technology of China, Hefei 230026, China}
\affiliation{School of Astronomy and Space Sciences, University of Science and Technology of China, Hefei 230026, China}
\email{}

\author[0000-0003-0694-8946]{Qingfeng Zhu}
\affiliation{Department of Astronomy, University of Science and Technology of China, Hefei 230026, China}
\affiliation{School of Astronomy and Space Sciences, University of Science and Technology of China, Hefei 230026, China}
\affiliation{Institute of Deep Space Sciences, Deep Space Exploration Laboratory, Hefei 230026, China}
\email{}

\author[0000-0003-3728-9912]{Xianzhong Zheng}
\affiliation{Tsung-Dao Lee Institute and Key Laboratory for Particle Physics, Astrophysics and Cosmology, Ministry of Education, Shanghai Jiao Tong University, Shanghai 201210, China}
\email{}

\begin{abstract}
The gravitational wave (GW) event S250206dm, as the first well-localized neutron star merger candidate potentially located in the mass gap, presented a unique opportunity to probe the electromagnetic signatures from such a system. Here we report a deep, multiband search with the new 2.5-meter Wide Field Survey Telescope (WFST), covering $\sim$64\% of the localization region up to a $5\sigma$ limiting magnitude of 23 mag. In total, 12 potential candidates have been identified while none of them are likely related to S250206dm. This non-detection provides the most stringent constraint to date on any associated kilonova. Crucially, an AT 2017gfo-like event at 269 Mpc can be excluded only by WFST observations. Based on ejecta mass limits, a neutron star–black hole with a large mass ratio ($Q\gtrsim3.2$) is disfavored. This optical-derived constraint on the mass ratio reaches, for the first time, a precision comparable to that inferred from the GW signal. This work presents the best observation of this type of events until now, and demonstrates the power of rapid, deep follow-up observations to constrain the properties of compact binary progenitors, offering key insights into the constituents of the mass gap.
\end{abstract}



\section{Introduction}
Binary neutron star (BNS) and neutron star--black hole (NSBH) mergers are gravitational wave (GW) sources, which can produce electromagnetic (EM) counterparts \citep{Abbott_2017_2,Metzger_2020,NAKAR20201}. As the first confirmed BNS merger, the GW event GW170817 \citep{Abbott_2017b}, and the subsequent discovery of its EM counterparts marked a breakthrough, opening the era of multi-messenger astronomy. Various EM counterparts to GW170817 were observed across the entire EM spectrum by numerous facilities. Approximately two seconds post-merger, the Fermi Gamma-ray Space Telescope detected the short gamma-ray burst (sGRB) GRB170817A, which lasted about two seconds \citep{goldstein_ordinary_2017,savchenko_integral_2017}. Subsequently, ultraviolet, optical, and near-infrared emissions were detected and confirmed as being from a kilonova (KN) coincident with GW170817 and GRB170817A, designated as AT2017gfo \citep{coulter_swope_2017,pian_spectroscopic_2017,tanvir_emergence_2017,shappee_early_2017}. The afterglow emissions were also detected later from X-ray to radio \citep{Troja_2017,DAvanzo_2018}. The EM counterpart observations, combined with GW data, can be applied to a wide range of scientific studies, such as the origin of heavy elements and constraints on the Hubble constant \citep{kasen_origin_2017,Abbott_2017Natur}.

KN is an approximately thermal transient powered by the radioactive decay of $r$-process nuclei, which has long been suspected to be generated in BNS or NSBH mergers \citep{lattimer1974,lattimer1977,eichler1989}. The case of a BNS merger being the progenitor of a KN was confirmed by the multi-messenger detection of GW170817. To search for the KNe from NSBH mergers, many observations were conducted to follow up NSBH merger candidates during the third observing run (O3) of the LIGO Scientific and Virgo Collaboration (LVC) \citep{Ackley_2020,Anand_2020}. However, no evident KN from a NSBH merger has been found so far. The non-detection results can be attributed to several main reasons. Firstly, whether tidal disruption occurs in a NSBH merger is crucial for producing a KN. Secondly, KN is a faint and fast-evolving transient, requiring a timely and deep follow-up to search. Finally, due to the large localization area, it is difficult to cover the entire skymap even for wide-field telescopes, and searching for KN candidates is also challenging considering the numerous transients in survey data.

During O4, the LIGO/Virgo/KAGRA (LVK) Collaboration has detected a few GW events likely originating from BNS or NSBH mergers with low false alarm rates (FARs). For NSBH merger candidates, only two events, GW230529\_181500 and S250206dm, have a significant probability of retaining neutron star material outside the merger remnant, potentially producing an EM counterpart \citep{Abac_2024,2025GCN.39231....1L}. Additionally, the source classifications of both events are uncertain. They may originate from either a BNS or a NSBH merger, with a high probability of involving one compact object in the so-called lower ``mass gap'' between the anticipated mass range ($\sim3$ to $\sim5\,M_\odot$) of NSs and that of BHs. The detection of an EM counterpart is an effective way to infer the nature of the composition of the merging system \citep{Barbieri_2019a,Kawaguchi_2020}. For GW230529\_181500, searching for its EM counterpart is challenging due to its poor localization, spanning a sky area of approximately 24,100 deg$^2$ at the 90\% credible level \citep{Abac_2024}. S250206dm, detected on 6 February 2025, was localized to a 90\% credible area of 547 deg$^2$ \citep{2025GCN.39231....1L}. The relatively precise localization makes the event a rare opportunity to search for an EM counterpart of a NSBH merger. \par

In this paper, we present the follow-up campaign conducted by the 2.5-meter Wide Field Survey Telescope (WFST) for S250206dm. In Section \ref{sec_2}, we introduce the GW event S250206dm and the follow-up observations by WFST. In Section \ref{sec_3}, the procedures of data reduction and EM counterpart search are presented. The probability of discovered candidates as the EM counterpart is discussed in detail in Section \ref{sec_4}. In Section \ref{sec_5}, based on the non-detection result, the constraints on the potential KN and the progenitor of S250206dm are presented. The comparison with other follow-up campaigns and potential joint constraint are discussed in Section \ref{sec_6} Finally, we conclude our search campaign and constraint results in Section \ref{sec_7}. Throughout this study, we adopt a standard $\Lambda$CDM cosmology with parameters $H_0=67.7\,\text{km s}^{-1}\,\text{Mpc}^{-1}$, $\Omega_M=0.31$ and $\Omega_\Lambda=0.69$ \citep{Planck2018}. 

\section{S250206dm and WFST follow-up campaign} \label{sec_2}
The GW event S250206dm was detected by the LIGO-Livingston and LIGO-Hanford detectors at 21:25:30.439 UTC on February 6, 2025, when Virgo and KAGRA were offline\footnote{\href{https://online.igwn.org}{https://online.igwn.org}}. After several alerts updated \citep{2025GCN.39231....1L}, S250206dm has a low FAR of 1 per 25.01 years and may originate from a BNS merger or a NSBH merger, with probabilities of 37\% and 55\%, respectively. The localization of S250206dm was estimated to have a median luminosity distance of 373 Mpc and a 90\% skymap area of $547\,\text{deg}^2$. Other significant compact binary merger events detected in O4, which have at least one NS pre-merger, are listed in Table \ref{tab:O4_events}. The threshold of $P_\text{BNS} + P_\text{NSBH} > 0.2$ is adopted to select events with at least one NS. For events with higher $P_\text{BBH}$, the systems have larger chirp masses and heavier BHs, where the NS is unlikely to be tidally disrupted unless the BH has a high aligned spin. According to the preliminary online estimates, for NSBH merger candidates, GW230529\_181500 and S250206dm are the most likely to produce EM counterparts, with $P_\text{HasRemnant} > 0$ for the two events only. Given the more precise localization, S250206dm is more feasible for current wide-field survey facilities to follow up.
\begin{table}[htbp]
    \centering
    \begin{threeparttable}
    \footnotesize
    \renewcommand{\arraystretch}{1.}
    \setlength{\tabcolsep}{2pt}
    \caption{\noindent\textbf{Properties of the significant BNS or NSBH candidates discovered in O4.} The information of these candidates is from GRACEDB\tnote{1}, where the BNS or NSBH candidates are selected by $P_\text{BNS} + P_\text{NSBH} > 0.2$.} 
    \begin{tabular}{l*{8}{c}}
        \toprule 
        Event & FAR (/yr) & $P_\text{BNS}$ & $P_\text{NSBH}$ & $P_\text{BBH}$ & $P_\text{HasRemnant}$ & $A_{90}\,(\text{deg}^2)$ & $D_\text{L}\,(\text{Mpc})$\\
        \midrule
        S230518h & 0.10 & 0\% & 86\% & 4\% & 0\% & 460 & $204\pm57$ \\
        GW230529\_181500\tnote{2} & 0.0062 & 31\% & 62\% & 0\% & 12\% & 24534 & $197\pm62$ \\
        S230627c & 0.10 & 0\% & 49\% & 48\% & 0\% & 82 & $291\pm64$ \\
        S240910ci & 0.10 & 0\% & 31\% & 69\% & 0\% & 394 & $662\pm166$ \\
        S241109bn & $4.5\times10^{-4}$ & 0\% & 72\% & 28\% & 0\% & 10138 & $603\pm159$ \\
        S250206dm & 0.04 & 37\% & 55\% & 0\% & 30\% & 547 & $373\pm104$ \\
        S250818k & 2.15 & 29\% & 0\% & 0\% & 100\% & 949 & $237\pm62$ \\
        \bottomrule
    \end{tabular}
    \begin{tablenotes}
        \footnotesize
        \item[1] \href{https://gracedb.ligo.org/}{https://gracedb.ligo.org/}
        \item[2] Although the data of GW230529\_181500 has been released, the online results of the event property is adopted here to compare with other events.
    \end{tablenotes}
    \label{tab:O4_events}
    \end{threeparttable}
\end{table}

We conducted a follow-up campaign for S250206dm using WFST, installed at the summit of the Saishiteng Mountain near Lenghu \citep{Deng_2021}, Qinghai province, China. WFST has a field of view (FoV) of 6.55 deg$^{2}$ and a 2.5-meter primary mirror, making WFST one of the most powerful transient survey facilities in the world \citep{Wang_2023,conroy2023china}. Extragalactic transients are primary targets for WFST \citep{Wang_2023}, including GW EM counterparts, supernovae, tidal disruption events, etc. WFST got the first light in September 2023, a few months after the start of O4. To search for potential EM counterparts of GW events detected in O4, a pre-research and delicate follow-up strategy through WFST target of opportunity (ToO) observations were developed in \cite{Liu_2023}. The ToO strategy was subsequently applied to observations of GW events S240422ed and S250206dm. 

\begin{figure}[htbp]
    \centering
    \begin{overpic}[width=1\textwidth]{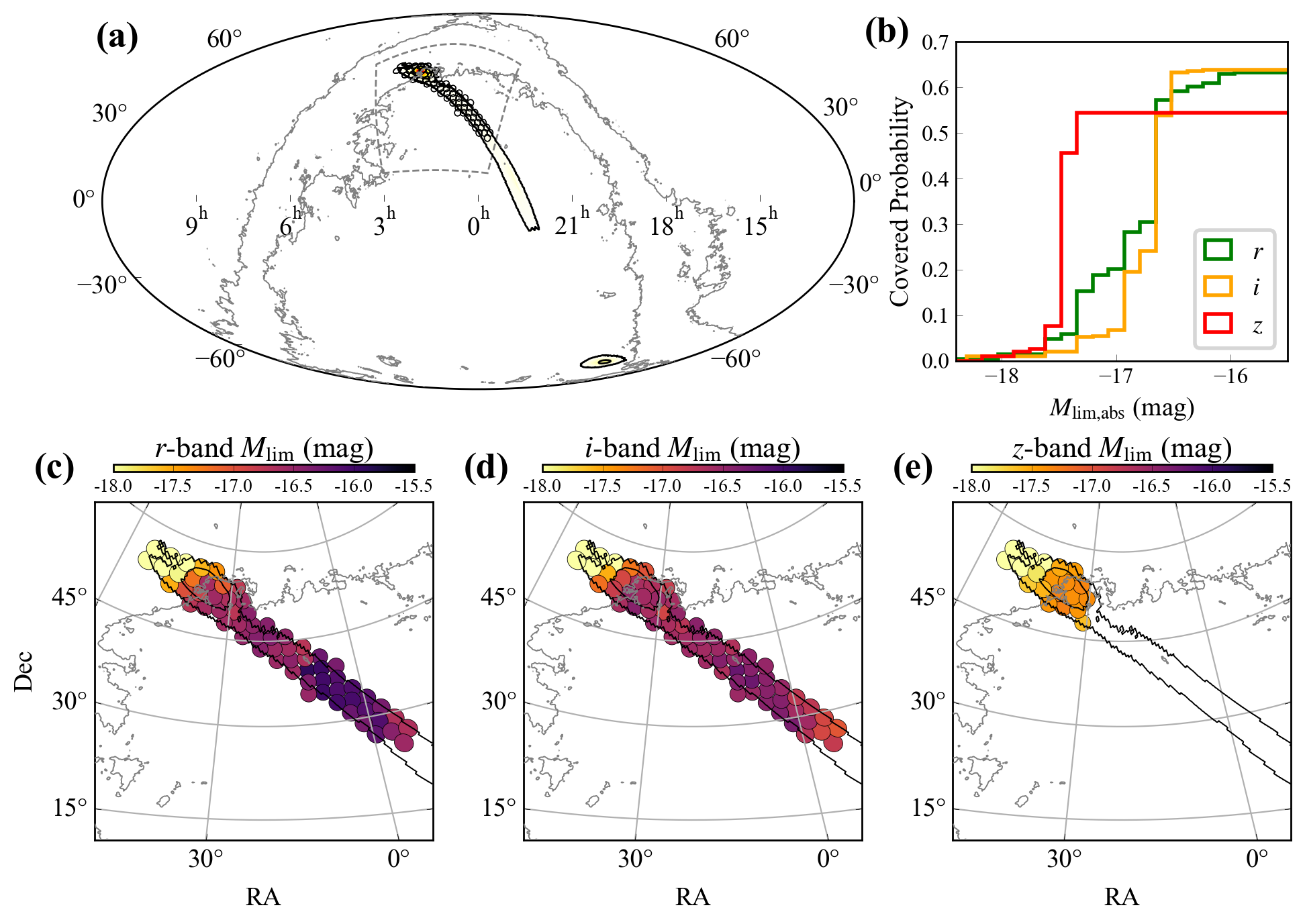}
    \end{overpic}
    \caption{Skymap coverage and multiband depths of WFST on the first night for S250206dm. Panel (a): the total coverage by WFST for S250206dm, where the Bilby skymap of S250206dm from \cite{2025GCN.39231....1L} is adopted, with 50\% and 90\% probability region contours. Galactic extinction is also considered, with the contour of $E(B-V)=0.3$ for reference \citep{Schlafly_2011}. Panel (b): The probability coverage by the WFST as a function of limiting absolute magnitudes in $r$, $i$, and $z$ bands.
    Panel (c), (d), and (e): The zoomed-in covered regions by WFST in $r$, $i$, and $z$ bands, respectively, corresponding to the region delineated by the dashed line in Panel (a). The color map represents the 5$\sigma$ limiting absolute magnitudes after stacking, converted based on the distance estimate for each WFST pointing.} 
    \label{fig:coverage}
\end{figure}

\begin{figure}[htbp]
    \centering
    \begin{overpic}[width=0.7\textwidth]{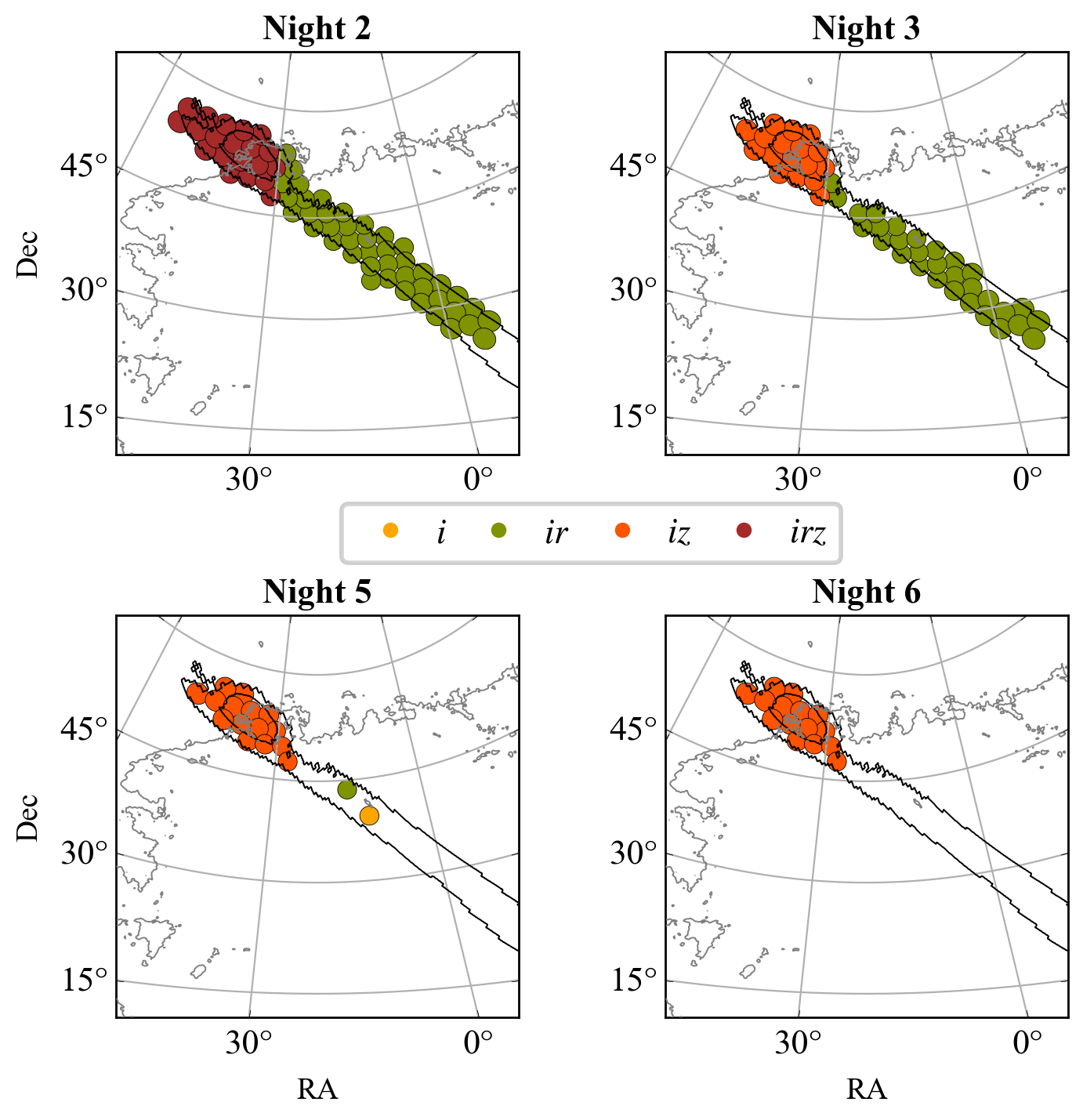}
    \end{overpic}
    \caption{Skymap coverage of WFST observations for S250206dm after the first night. The Bilby skymap of S250206dm from \cite{2025GCN.39231....1L} is adopted, with 50\% and 90\% probability region contours shown. Galactic foreground is shown as a contour of $E(B-V)=0.3$ \citep{Schlafly_2011}.}
    \label{fig:coverage_other}
\end{figure}

\begin{table}[htbp]
    \centering
    \footnotesize
    \begin{threeparttable}
        \renewcommand{\arraystretch}{1.5}
        \setlength{\tabcolsep}{3pt}
        \caption{\noindent\textbf{The summary of WFST follow-up observations for S250206dm.}} 
            \begin{tabular}{*{6}{c}}
                \toprule 
                Night & Start time (UTC)    & End time (UTC)      & Band  & Area\tnote{1} (deg$^2$) & $P_\text{covered}$\tnote{1}  \\
                \midrule
                1     & 2025-02-07T12:10:10 & 2025-02-07T17:01:21 &$r,i,z$& 345   & $\sim64$\%          \\
                2     & 2025-02-08T12:14:53 & 2025-02-08T17:15:32 &$r,i,z$& 345   & $\sim64$\%          \\
                3     & 2025-02-09T12:14:04 & 2025-02-09T17:00:05 &$r,i,z$& 259   & $\sim63$\%          \\
                4     & \multicolumn{5}{c}{No observation due to weather}                                 \\
                5     & 2025-02-11T13:00:10 & 2025-02-11T16:49:24 &$i,z$  & 97    & $\sim55$\%          \\
                6     & 2025-02-12T13:28:49 & 2025-02-12T17:06:47 &$i,z$  & 85    & $\sim54$\%          \\
                \bottomrule
            \end{tabular}
        \begin{tablenotes}
            \footnotesize
            \item[1] Corresponding to total covered area and probability of all bands, where the FoV of WFST is approximated as a circle with $1.4^\circ$ radius.
        \end{tablenotes}
        \label{tab:obs_summary}
    \end{threeparttable}
\end{table}

For S250206dm, WFST observations started approximately 14.7 hours (12:10 UTC on February 7, 2025) after the GW detection and lasted for a week \citep{2025GCN.39249....1L,2025GCN.39293....1X}, primarily focused on the northern part of skymap. To mitigate the effects of moonlight (moon phase of $\sim0.7$, increasing in subsequent nights) and Galactic extinction, a longer-wavelength combination of $r$, $i$, and $z$ bands was adopted. Additionally, given the high density of stars in the low Galactic latitude region of S250206dm skymap, exposure times of $1\times90$ s and $2\times60$ s for each pointing each band were adopted to prevent excessive saturation. For pointings with multiple coverages per night, they were dithered to fill the CCD gap.

The WFST coverage and limiting magnitudes on the first night are shown in Figure \ref{fig:coverage}. To evaluate the search capability, the limiting magnitudes in each band are converted to absolute magnitudes using the median distance estimate from the GW skymap in Panel (c), (d), and (e). The effect of Galactic foreground extinction is also included. The host extinction is not considered in this work due to the generally large projected distance from galactic center \citep{Fong_2022}. The extinction correction in this work is based on $E(B-V)$ map from \cite{Schlafly_2011}, with $R_V = 3.1$. For KN AT 2017gfo, the peak absolute magnitudes are approximately $-16$ mag in $r$, $i$, and $z$ bands \citep{Abbott_2017_2}. With a limiting magnitude range from $-15.8$ mag to $-18.0$ mag approximately, our observations achieve the depth required to detect an AT 2017gfo-like KN at the peak phase. 

The observing strategies for the second and third nights were similar to that of the first night. As the potential KN became fainter and the localization was updated, a strategy with a longer exposure time and a smaller coverage area was adopted. Specifically, for the fifth and the sixth nights, the reduced coverage region is shown in Figure \ref{fig:coverage_other}, where $i$ and $z$ bands with exposure times of $3\times120$ s and $2\times180$ s were used, respectively. The observation on the fourth night was not conducted due to snowy weather at the summit. The covered area, probability, and observation time range for each night are summarized in Table \ref{tab:obs_summary}. In total, $\sim64\%$ and $\sim55\%$ Bilby skymap \citep{2025GCN.39231....1L} were covered by WFST in the first three nights and the last two nights, respectively.

\section{Search for EM counterpart} \label{sec_3}
The raw images from the follow-up observations of S250206dm were processed using the WFST pipeline, which is mainly based on the the Vera C. Rubin observatory (VRO/LSST) image processing pipeline (hereafter the ``LSST pipeline'') \citep{boschOverviewLSSTImage2018}. The LSST pipeline is a modular and efficient software suite written in a combination of Python and C\texttt{++} to process massive survey data. The main processing steps of the LSST pipeline include data ingestion, instrument signature removal, point spread function (PSF) modeling, astrometric and photometric calibration, source detection and deblending, image alignment and coaddition, image subtraction, and alert product generation. The WFST pipeline was modified from the LSST pipeline in two main aspects: the addition of extension files defining the CCD configuration and the replacement of the subtraction software with \texttt{SFFT} to optimize and expedite the subtraction process \citep{Hu_2022}. Please see \cite{Cai_2025} for details of the WFST pipeline. \par

\begin{table}[htbp]
    \centering
    \footnotesize
    \renewcommand{\arraystretch}{1.}
    \setlength{\tabcolsep}{2pt}
    \caption{\noindent\textbf{The number of alert after series of filters for candidate search.}} 
    \begin{tabular}{l*{2}{c}}
        \toprule 
        Description           & Filter                      & Alert number      \\
        \midrule
        Total                 &                             & 3746535           \\
        Multiple detections   & $N_\text{det}\geq2$         & 1735634           \\
        Real source           & $S_\text{RB}\geq0.1$        & 93180             \\
        Positive source       & $N_\text{negative}=0$       & 48771             \\
        Not variable star     & $d_\text{point}\geq1''$     & 5002              \\
Far from bright source        & $d_\text{15mag}\geq10'', d_\text{12mag}\geq40''$ & 2236 \\
Not galactic nucleus activity & $d_\text{nuc}\geq0.5''$     & 2172               \\
        Not moving object     & $d_\text{SkyBot}\geq3''$    & 2172               \\
        \bottomrule
    \end{tabular}
    \label{tab:alert_filter}
\end{table}

To search for the EM counterpart timely, although the lack of WFST archival images for the S250206dm skymap, the preliminary search was conducted using the first night data as reference images, but no valid candidate was found. To ensure that the potential counterpart is faint enough, reference images were taken about two weeks later (2025-02-16T12:44 to 2025-02-21T14:59 UTC), millions of alerts were detected through basic image reduction, stacking for the images in the same band each night, and subtraction using the WFST pipeline. We searched for the KN using the following filtering criterias to exclude the contamination:
\begin{itemize}
    \item[1.]Not bogus or artifacts: Require multiple detections of $N_\text{det}\geq2$ and $S_\text{RB}\geq0.1$, where $S_\text{RB}$, calculated by a real-bogus (RB) classifier \citep{Liu_2025}, represents the fraction of real detections for an alert.
    \item[2.]Not variable stars: Exclude alerts close to known point sources.
    \item[3.]Not close to bright sources: Exclude alerts around bright sources ($M<15$ mag), which may produce false detections due to saturation.
    \item[4.]Not galactic nucleus activity: Exclude alerts close to centers of known galaxies.
    \item[5.]Not moving object: Exclude alerts close to known solar system objects.
\end{itemize}
The number of alerts remained after each filter is shown in Table \ref{tab:alert_filter}, where crossmatches of variable stars, bright sources and galactic centers are preformed using Pan-STARRS1 (PS1) DR2 and Gaia DR3 catalogs \citep{Chambers_2016,Magnier_2020,GaiaCollaboration_2023}. Point and extended sources are distinguished by star classification probability and the difference between results of PSF and Kron photometry (with sources satisfying $M_{i,\text{PSF}} - M_{i,\text{Kron}} < 0.05$ classified as stars\footnote{\href{https://outerspace.stsci.edu/display/PANSTARRS/How+to+separate+stars+and+galaxies}{https://outerspace.stsci.edu/display/PANSTARRS/How+to+separate+stars+and+galaxies}}) for the Gaia and PS1 databases, respectively. To exclude moving objects, the remaining alerts were crossmatched with Sky Body Tracker to rule out Solar System objects \citep{SkyBot_2006}. Finally, 2172 alerts remained after applying an inclusive set of filters to avoid missing potential candidate. These remaining alerts will undergo visual inspection based on analogous criteria described in Table \ref{tab:alert_filter} to exclude artifacts and other contamination.

\begin{sidewaystable}
    \centering
    \begin{threeparttable}
    \footnotesize
    \renewcommand{\arraystretch}{1.5}
    \setlength{\tabcolsep}{5pt}
    \caption{\noindent\textbf{Properties of 12 candidates after human vetting and specific reasons for exclusion as a KN counterpart.}} 
    \begin{tabular}{*{12}{c}}
        \toprule 
        No.
        &$N_\text{det}$\tnote{1}
        &Ra
        &Dec
        &$\dot{M_i}$\tnote{2}
        &Band
        &$M_{i\text{,p}}$\tnote{2}
        &$M^\prime_{i\text{,p}}$\tnote{3}
        &$D_\text{L,host}$\tnote{4}
        &$\text{CDF}(D_\text{L})$\tnote{5}
        &Offset
        &Reason\\
              &   &          &          &(mag/day)&      & (mag) & (mag) &(Mpc)&      &(Kpc)&                    \\
        \midrule
        1 & 7 & 40.01920 & 50.88298 & $0.08\pm0.03$  & $r,i,z$& $-16.7\pm0.5$ & $-17.4\pm0.5$ & $395\pm75^\text{b}$& 0.63 & 7.5 & Previous detection \\
        2 & 4 & 356.20686& 28.06492 & $0.22\pm0.11$  & $r,i$  & $-19.2\pm0.1$ & $-19.4\pm0.1$ & 446 & 0.54 & 5.5 & Previous detection \\
        3 & 4 & 35.91705 & 53.46289 & $0.02\pm0.03$  & $i$    & $-18.6\pm0.3$ & $-19.0\pm0.3$ & $780\pm81^\text{a}$ & 0.94 & 15.2& Skymap mismatched  \\
        4 & 4 & 30.12691 & 48.66882 & $0.00\pm0.03$  & $i$    & $-18.3\pm0.7$ & $-18.7\pm0.7$ & $705\pm203^\text{a}$ & 0.99 & 2.7 & Slow evolution \\
        5 & 4 & 29.93754 & 51.35800 & $-0.02\pm0.03$ & $i$    & $-18.1\pm0.7$ & $-18.5\pm0.7$ & $589\pm193^\text{a}$ & 0.98 & 4.1 & Slow evolution \\ 
        6 & 3 & 30.88382 & 49.36016 & $0.14\pm0.07$  & $i$    & $-19.1\pm0.7$ & $-19.6\pm0.7$ & $990\pm225^\text{a}$ & 1.00 & 4.5 & Too bright \\
        7 & 3 & 38.80782 & 56.26273 & $-0.05\pm0.05$ & $i$    & $-17.3\pm0.4$ & $-18.5\pm0.4$ & $473\pm76^\text{a}$ & 0.86 & 7.1 & Slow evolution \\
        8 & 2 & 3.45112  & 32.29102 & $0.03\pm0.11$  & $i$    & $-17.1\pm0.4$ & $-17.2\pm0.4$ & $471\pm76^\text{a}$ & 0.62 & 4.7 & Too bright? \\
        9 & 2 & 4.47506 & 32.98365 &$-0.02\pm0.12$ & $i$   & $-17.3\pm0.5$ & $-17.4\pm0.5$ & $472\pm76^\text{a}$ & 0.64 & 1.2 & Too bright? \\
        10 & 2 & 40.15301& 53.50889 & $0.02\pm0.03$ & $i$   & $-20.5\pm0.6$ & $-21.4\pm0.6$ &$2158\pm504^\text{b}$& 1.00 & 3.5 & Skymap mismatched  \\
        11 & 3 & 32.21519& 53.05843 & $0.03\pm0.03$ & $i$   & $-20.6\pm0.5$ & $-21.0\pm0.5$ &$1789\pm377^\text{b}$& 0.99 & 3.4 & Skymap mismatched  \\
        12 & 12 & 35.70140&	50.28884 & $0.03\pm0.01$ &$r,i,z$& $-20.4\pm0.6$ & $-21.05\pm0.6$ &$1727\pm375^\text{b}$& 0.99 & 4.4 & Skymap mismatched  \\
        \bottomrule
    \end{tabular}
    \begin{tablenotes}
        \footnotesize
        \item[1] Total number of detections of each candidate, after stacking of images with same band each night. All uncertainties are given at the 1$\sigma$ level.
        \item[2] The magnitude changing rate and peak absolute magnitude of each candidate are listed in $i$ band. The error of peak absolute magnitudes include the uncertainties of the host distances.
        \item[3] The peak absolute magnitude of each candidate after Galactic extinction correction.
        \item[4] The spectroscopic redshift result of No. 2 candidate is from DESI DR1 \citep{desicollaboration2025datarelease1dark}. Distance with marker `a' is based on the photometric redshifts collected in GLADE+ catalog \citep{Dalya_2022}. If there is no match with GLADE+ catalog, the host distance is based on photometric redshift result of PS1 survey \citep{Beck_2021}, and is marked with `b'.
        \item[5] The cumulative conditional probability given the host distance and the candidate 2D location, based on the Bilby skymap. For candidates excluded by ``Skymap mismatched'', the host distance of $\mu-3\sigma$ is adopted to show the deviation considering uncertainty of photometric redshifts.
    \end{tablenotes}
    \label{tab:candidates}
    \end{threeparttable}
\end{sidewaystable}

\section{Candidate analysis} \label{sec_4}
After visual inspection for the filtered alerts, 12 candidates were identified as off-nucleus extragalactic transients, as listed in Table \ref{tab:candidates}. To exclude transients occurred before the GW detection, these candidates were crossmatched with public survey databases, including Transient Name Server \citep[TNS;][]{TNS}, Zwicky Transient Facility \citep[ZTF;][]{Bellm_2019,graham_zwicky_2019} ALeRCE explorer \citep{Forster_2021}, and the Asteroid Terrestrial-impact Last Alert System \citep[ATLAS;][]{Tonry_2018,Smith_2020} forced photometry server \citep{ATLAS_force}. Two candidates are excluded as potential counterparts due to archival detections by ZTF. Additionally, assuming the nearest galaxy as the host galaxy, the luminosity distances of remaining 10 candidates were obtained from the galaxy redshifts by crossmatching with the GLADE+ catalogs and PS1-based photometric redshift catalogs \citep{Beck_2021,Dalya_2022}. The percentiles of the marginal cumulative distribution function (CDF) at the locations of candidates are listed in Table \ref{tab:candidates}, with four candidates excluded due to significantly larger distances compared with GW localization even when uncertainties of photometric redshifts are taken into account. For the 12th candidate, the subtraction was re-preformed with the archival images of PS1 $3\pi$ survey as reference \citep{Beck_2021}, due to that it is still visible in the WFST template images. 

\begin{figure}[htbp]
    \centering
    \begin{overpic}[width=0.35\textwidth]{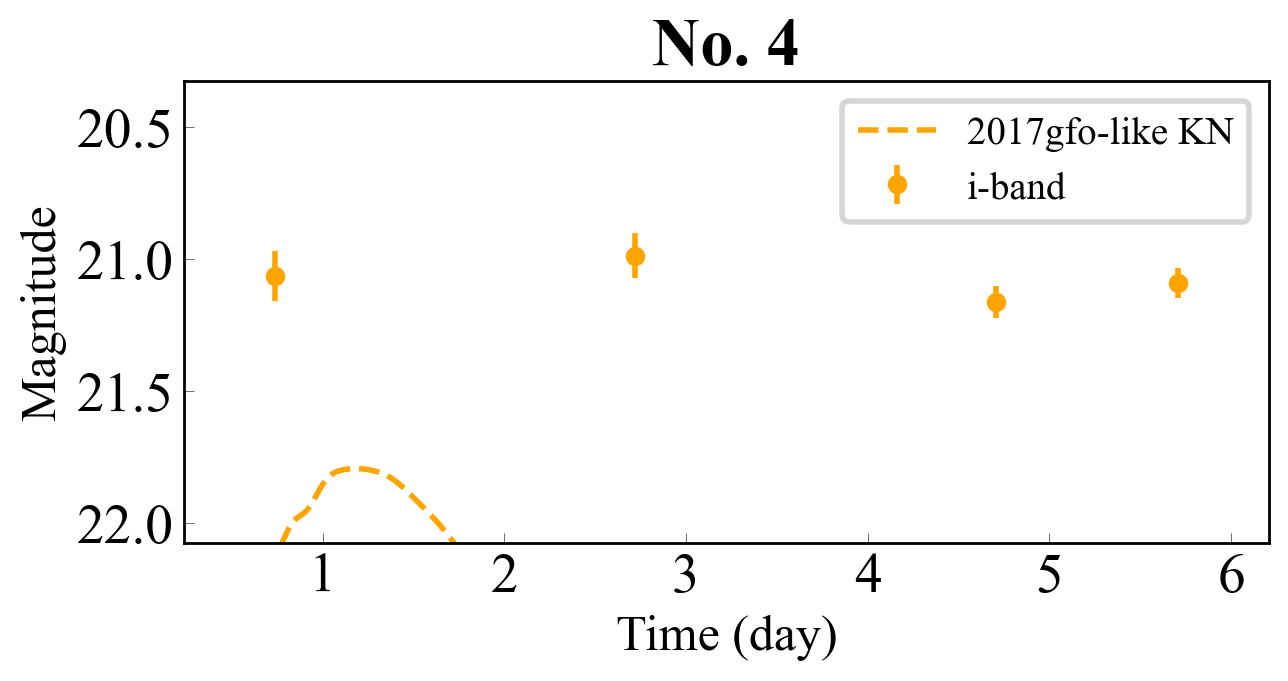}
    \end{overpic}
    \begin{overpic}[width=0.35\textwidth]{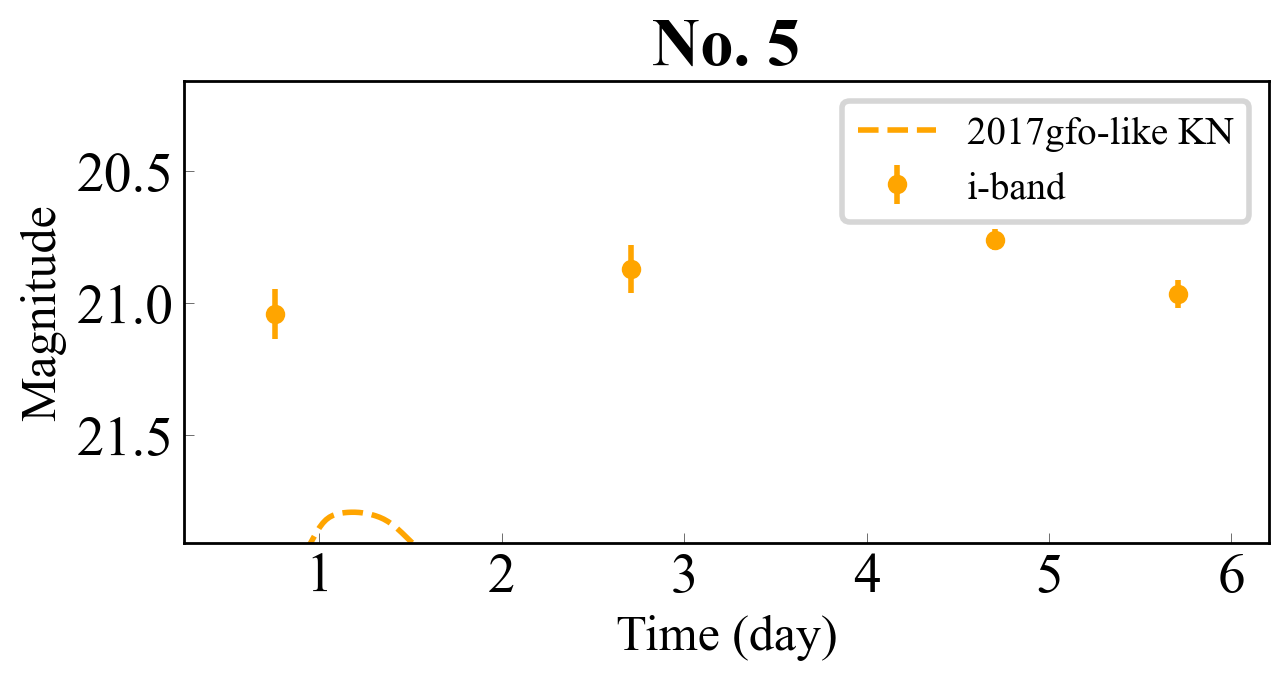}
    \end{overpic}
    \begin{overpic}[width=0.35\textwidth]{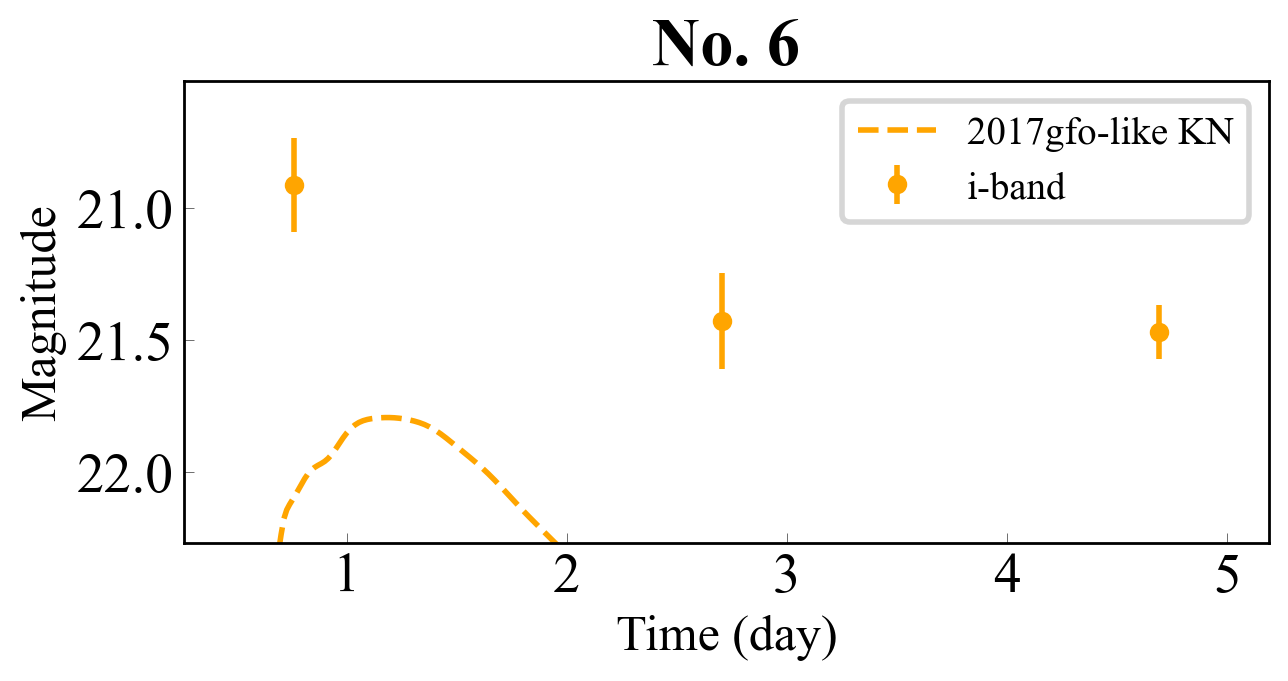}
    \end{overpic}
    \begin{overpic}[width=0.35\textwidth]{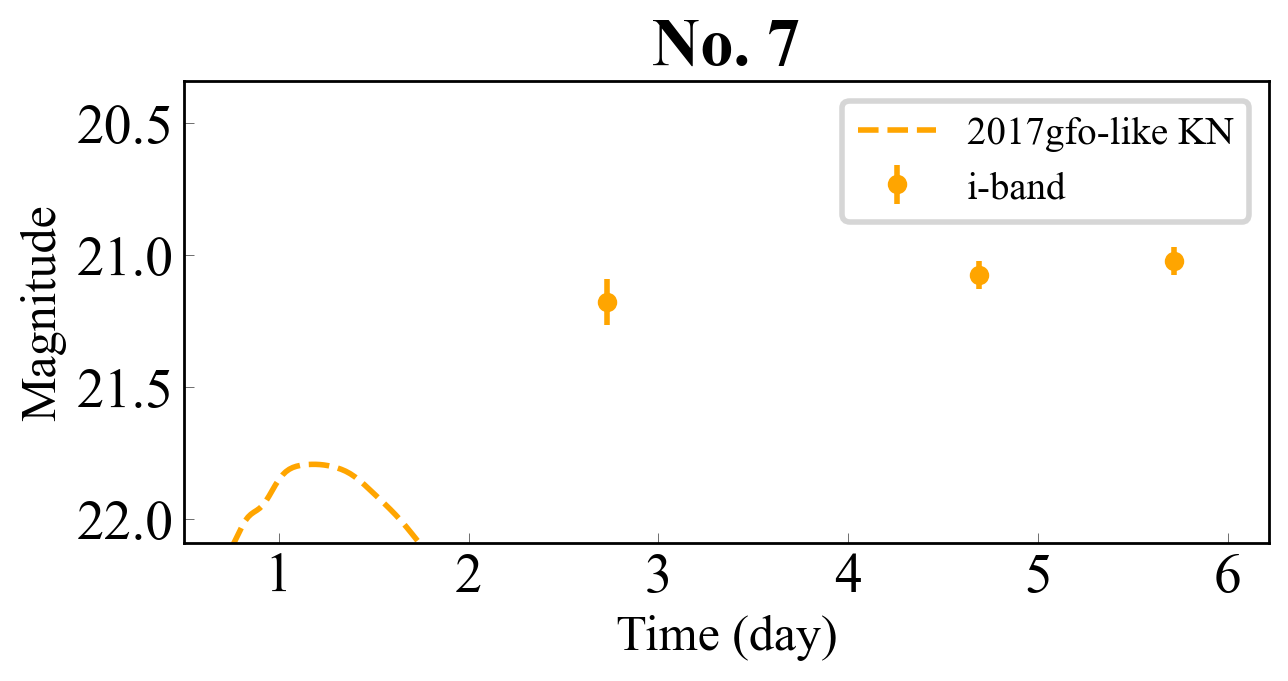}
    \end{overpic}
    \begin{overpic}[width=0.35\textwidth]{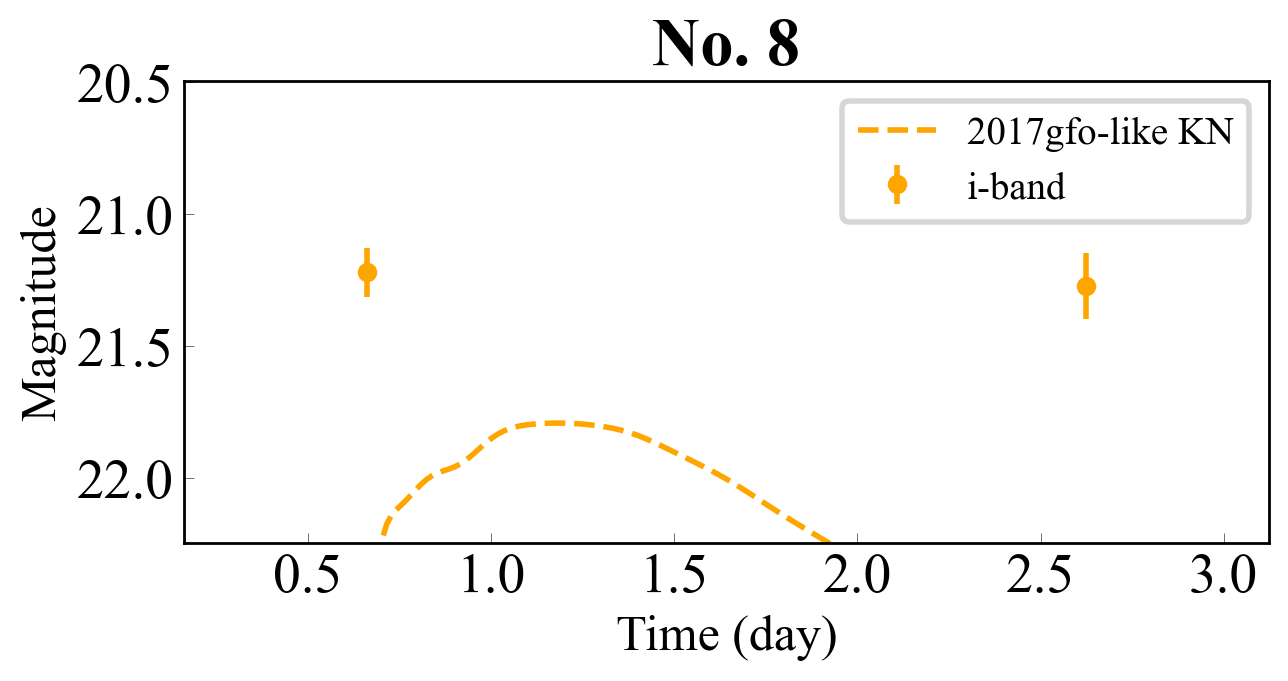}
    \end{overpic}
    \begin{overpic}[width=0.35\textwidth]{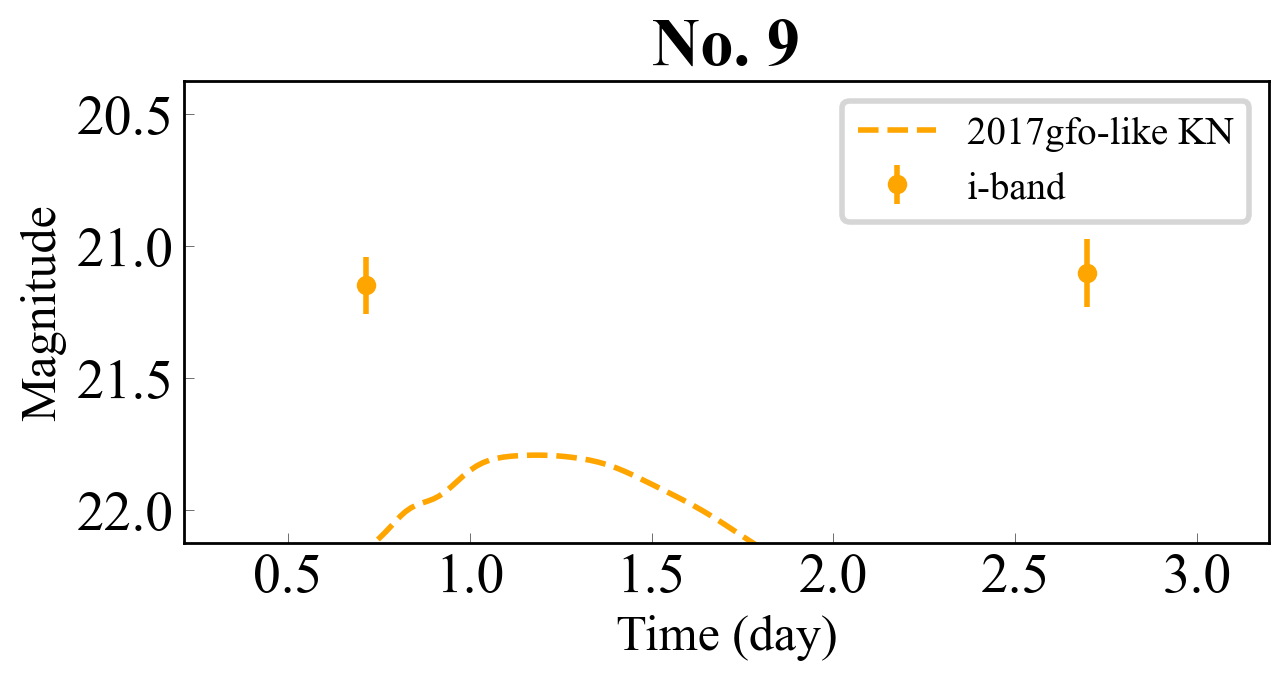}
    \end{overpic}
    \caption{Lightcurves of the six WFST candidates tagged by ``Too bright'' or ``Slow evolution''. The luminosity distance of a 2017gfo-like KN is set as the median of 373 Mpc.}
    \label{fig:LC6}
\end{figure}

\begin{figure}[htbp]
    \centering
    \begin{overpic}[width=0.4\textwidth]{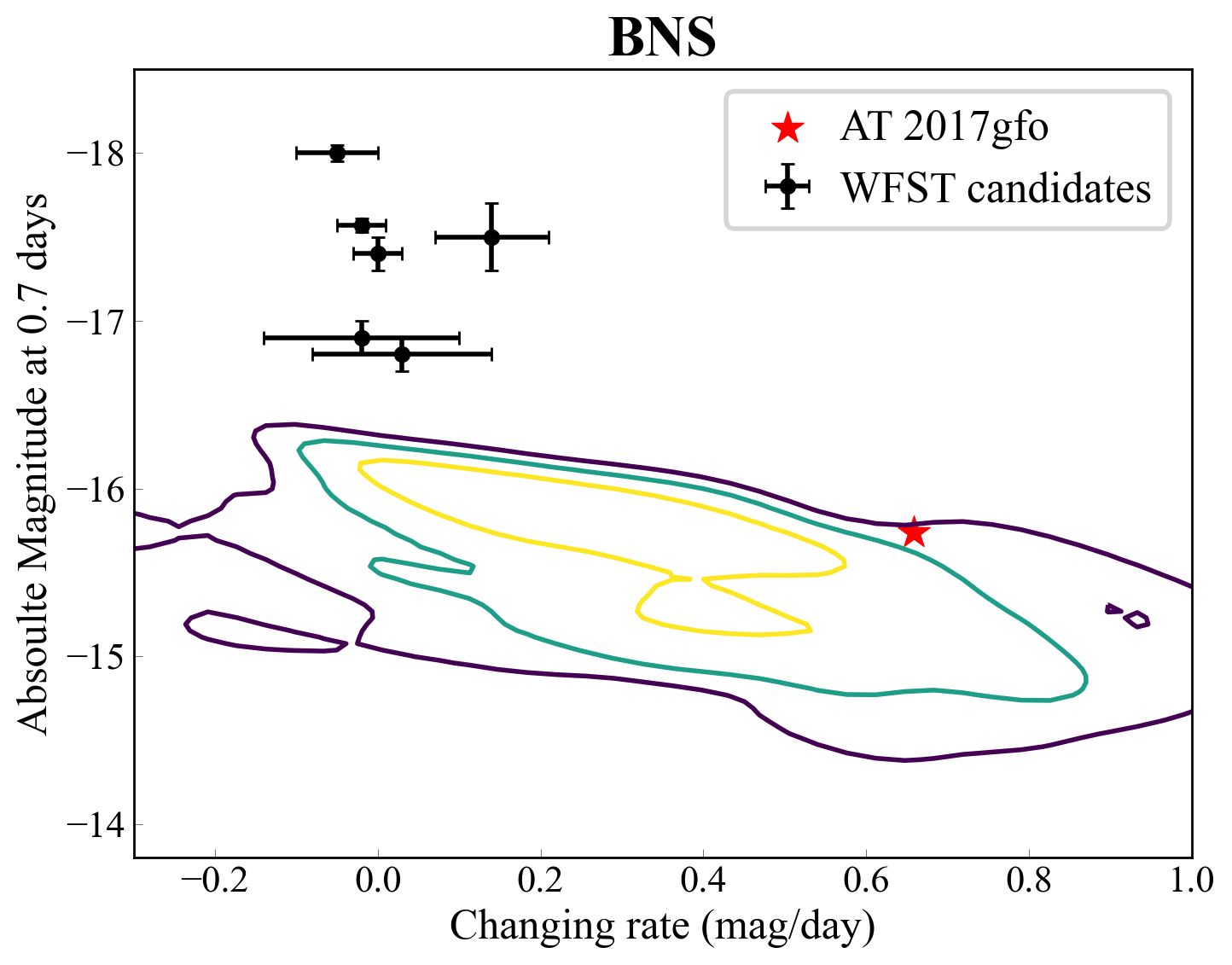}
    \put(0,72){{\textbf{(a)}}}
    \end{overpic}
    \begin{overpic}[width=0.4\textwidth]{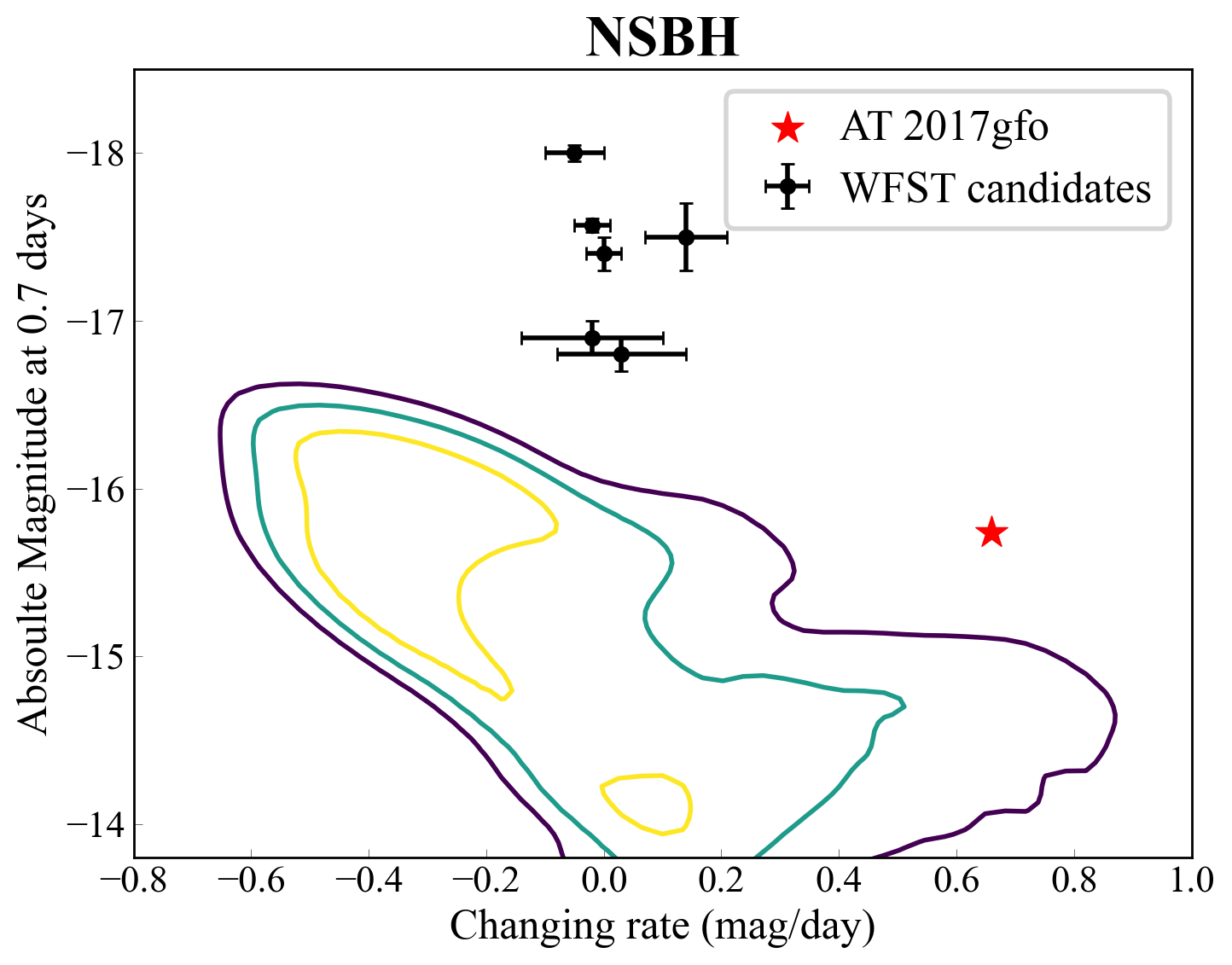}
    \put(0,72){{\textbf{(b)}}}
    \end{overpic}
    \caption{The model-dependent space of peak magnitude and variability in $i$ band. Panel (a) and (b): The KN model \texttt{POSSIS} is adopted for simulating KNe in the BNS and NSBH mergers, respectively. The black dots and red pentacles represent the six candidates tagged by ``Too bright'' or ``Slow evolution'' and AT 2017gfo, respectively. To derive the absolute magnitude, the luminosity distance is set as 373 Mpc for the candidates with no spectroscopic redshift available. Three contour lines correspond to 50\%, 90\%, 99\% percentiles of KN model regions. }
    \label{fig:KN_space}
\end{figure}

For the remained six candidates, we further compared their absolute magnitudes and variability (after Galactic extinction correction) with the KN model. The WFST lightcurves of the six candidates are shown in Figure \ref{fig:LC6}. The peak absolute magnitudes and evolution of these candidates are shown in Figure \ref{fig:KN_space}. The spaces of KNe based on the model \texttt{POSSIS} \citep{Bulla_2019,Anand_2020,Bulla_2023} are shown for comparison (details of the model please see Section \ref{sec_5}). The sampling times of simulated KN lightcurves used to derive absolute magnitudes and variability are $\sim0.7$ and $\sim2.7$ days post-merger, corresponding to the detection times of most candidates. Regarding their evolution, the decay rates of these six candidates are slower than those of KNe produced by BNS mergers but consistent with the range of decay rates for NSBH mergers. In terms of luminosity, all candidates are brighter than the luminosity range derived from the model, where the luminosity distance is set as 373 Mpc due to no spectroscopic redshift available for these candidates. Two candidates (the 8th and 9th) have peak magnitudes close to those of the simulated KN sample when the uncertainty of their photometric redshifts is considered. However, it remains challenging to classify them as KN candidates due to their limited number of detections. For other candidates, to produce such a bright KN, additional energy sources beyond radioactive decay are required, such as central engine activity \citep{Yu_2013}. Some magnetar-enhanced KN candidates associated with short gamma-ray bursts were discovered \citep{Gao_2017}, but no valid enhanced KN has been confirmed to date. Additionally, the merger remnant of S250206dm is unlikely to be a magnetar in the case of BNS merger due to its large chirp mass, which tends to collapse directly into a BH.  \par

To analyze the possibility as the afterglow counterpart of S250206dm for these six candidates, the scenarios of the BNS and NSBH merger are discussed as follows, respectively. For a BNS merger, the afterglow lightcurves with various viewing angles are simulated using the model \texttt{afterglowpy} \citep{Ryan_2020}, where the best-fitting results in \cite{Ryan_2020} of the GRB 170817A afterglow are adopted for other model parameters. Producing an afterglow with similar luminosity to the candidates requiring a small viewing angle ($\theta_\text{obs}<3^\circ$), but the evolution of such afterglow is quite fast compared with the candidate lightcurves. For a NSBH merger, if these candidates originated from the afterglow, their slow evolutions correspond to the afterglow peak phase. According to:
\begin{equation}
    t_{\text{peak},i}\simeq1.2\times10^3(1+z)\left(\frac{E_{\text{K},i}}{10^{53}\,\text{erg}}\right)^{(1/3)}\left(\frac{n}{1\,\text{cm}^{-3}}\right)^{(-1/3)}\left(\frac{\Gamma_i}{50}\right)^{(-8/3)}\,\text{s},
\end{equation}
$t_{\text{peak},i}\approx2$ days corresponds to a slow Lorentz factor ($\Gamma_i<5$), where $E_{\text{K},i} = 10^{55}\,\text{erg}$ and $n = 0.1\,\text{cm}^{-3}$ \citep{Zhang_2018}. However, such a slow Lorentz factor is inconsistent with a jet from a NSBH merger due to a limited baryon loaded. Therefore, these six candidates are unlikely related to the KN or afterglow counterpart of S250206dm. No spectral follow-up was carried out to further confirm the 12 candidates due to their faint brightnesses (mostly $>21$ mag) and reasons for exclusion. 

Candidates reported in the community were also checked in WFST data. These candidates in the optical band are from the ZTF team \citep{Ahumada_2025} and TNS\footnote{\href{https://www.wis-tns.org/ligo/o4/S250206dm_20250206_212530}{https://www.wis-tns.org/ligo/o4/S250206dm\_20250206\_212530}}. Candidates reported by the GW MultiMessenger Astronomy Dark Energy Camera \citep[DECam;][]{flaugher_dark_2015} Survey team \citep{Hu_2025} are not included due to limited overlap in the covered regions between WFST and Wendelstein. The coverage and detection of candidates by WFST are listed in Table \ref{tab:other_candidates_check} in the appendix, where candidates reported before the GW detection time and with Dec $\leq 0^\circ$ are not included. It is not surprising that most candidates were not detected in the WFST difference images due to the short time interval of approximately two weeks between the science and the reference images. Based on this time interval, the WFST follow-up observations are not sensitive to sources with slow evolution. For example, the candidate AT 2025bay, reported by ZTF, has a magnitude difference of $\sim0.2$ mag in $r$ band between the time of obtaining WFST science images and that of reference images, according to the lightcurve in \cite{Ahumada_2025}. This variability, corresponding to $\sim22.5$ mag in WFST difference images, is near the detection limit of WFST. Such sources with slow evolution are faint and difficult to be detected in the WFST observations. Based on WFST detections, no candidate is likely to be the KN counterpart. Additionally, in the X-ray band, candidates reported by the Einstein Probe through GCN Circulars \citep{2025GCN.39545....1L} were also checked in our data. However, no valid extragalactic source was discovered within the positional error of the candidates. Consequently, no robust KN candidate was discovered in the WFST follow-up observations. \par

\section{Counterpart constraint \& insights into origin} \label{sec_5}
Although non-detection of EM counterpart of GW events, the KN luminosity, ejecta mass during merger, and merger system property can be constrained by follow-up observations assuming a KN model \citep{Kasliwal_2020,Anand_2020,Ackley_2020,Pillas_2025}. A state-of-art KN model \texttt{POSSIS} is adopted in our analysis. \texttt{POSSIS} is a radiative transfer code for KNe developed by \cite{Bulla_2019}, including dynamical ejecta produced by collisions and tidal interactions, as well as disk wind ejecta generated by outflows from the post-merger accretion disk. For a BNS merger, post-merger ejecta and dynamical ejecta are distributed near the polar axis and the equatorial plane, respectively. As the grid model is based on simulations, the velocity ranges of dynamical and post-merger ejecta are fixed at 0.1--0.3 $c$ and 0.025--0.1 $c$, respectively. The free parameters of the model include dynamical ejecta mass $M_\text{dyn}$, post-merger ejecta mass $M_\text{pm}$, half-opening angle of the dynamical ejecta $\theta_\text{open}$ and viewing angle $\theta_\text{obs}$. A surrogate model from \cite{Lukosiute_2022} is adopted to obtain results for arbitrary parameters in our calculations.  

For AT 2017gfo-like KNe, the lightcurves are produced using \texttt{POSSIS} with the parameters from the fitted results in \cite{Dietrich_2020}: $M_\text{dyn} = 0.005\,M_\odot$, $M_\text{pm} = 0.052\,M_\odot$, $\theta_\text{open} = 49.5^\circ$, and $\theta_\text{obs} = 42.8^\circ$. For general BNS mergers, the ranges of dynamical ejecta $M_\text{dyn}$ and post-merger ejecta $M_\text{pm}$ are 0.0025 to 0.02 $M_\odot$ and 0.01 to 0.1 $M_\odot$, respectively. For NSBH mergers, the ranges of $M_\text{dyn}$ and $M_\text{pm}$ are both 0.01 to 0.09 $M_\odot$. For BNS and NSBH mergers, $\theta_\text{open}$ are fixed at $45^\circ$ and $30^\circ$ according to \cite{Bulla_2019} and \cite{Anand_2020}, respectively, and $\theta_\text{obs}$ are both sampled uniformly. Additionally, the distance and the viewing angle are randomly combined in our analysis due to the lack of further estimate from the GW signal.

\begin{figure}[t]
    \centering
    \makeatletter
    \begin{minipage}[b]{0.48\textwidth}
        \centering
        \begin{overpic}[width=\linewidth]{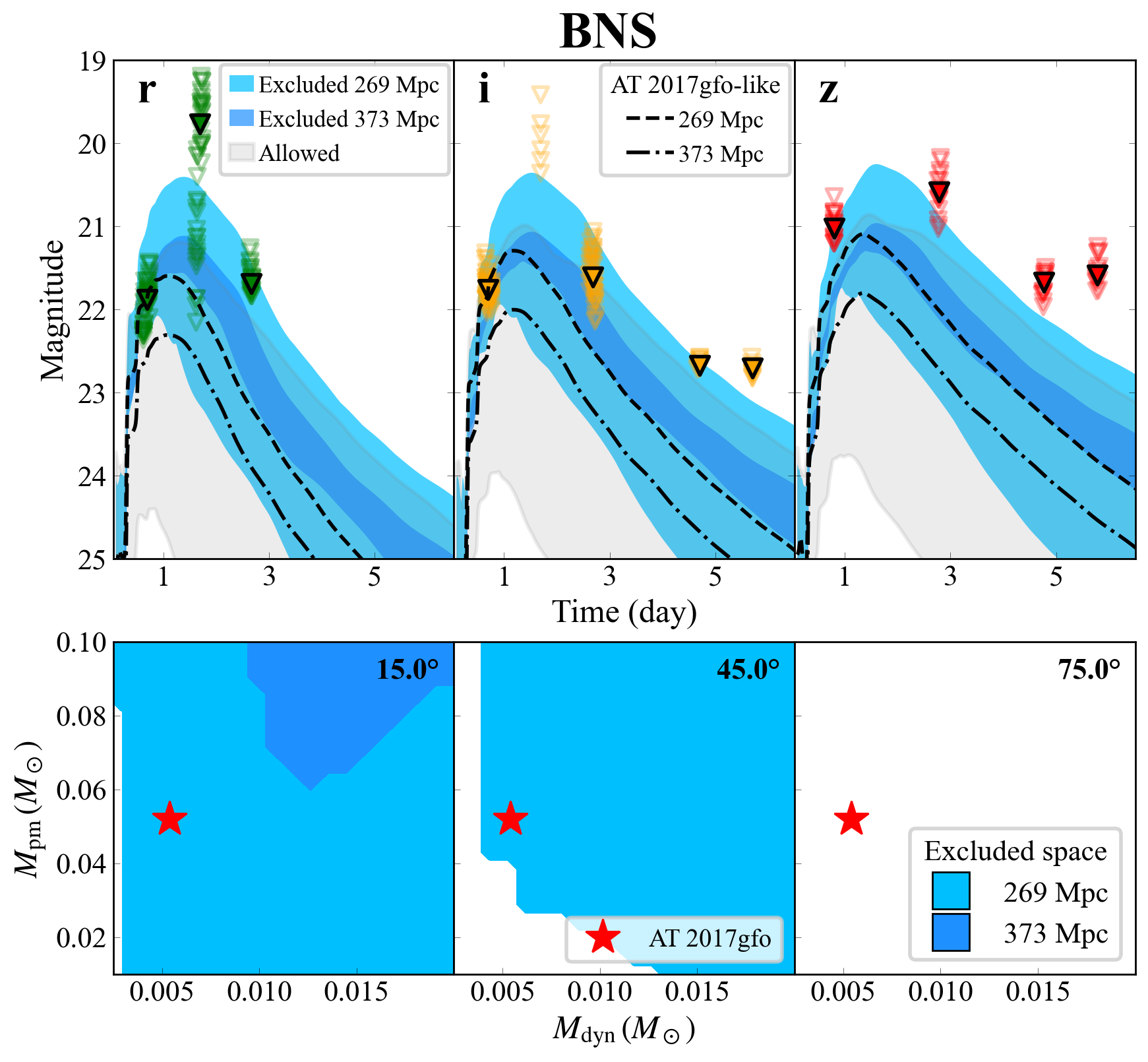}
        \end{overpic}
        
    \end{minipage}
    \begin{minipage}[b]{0.48\textwidth}
        \centering
        \begin{overpic}[width=\linewidth]{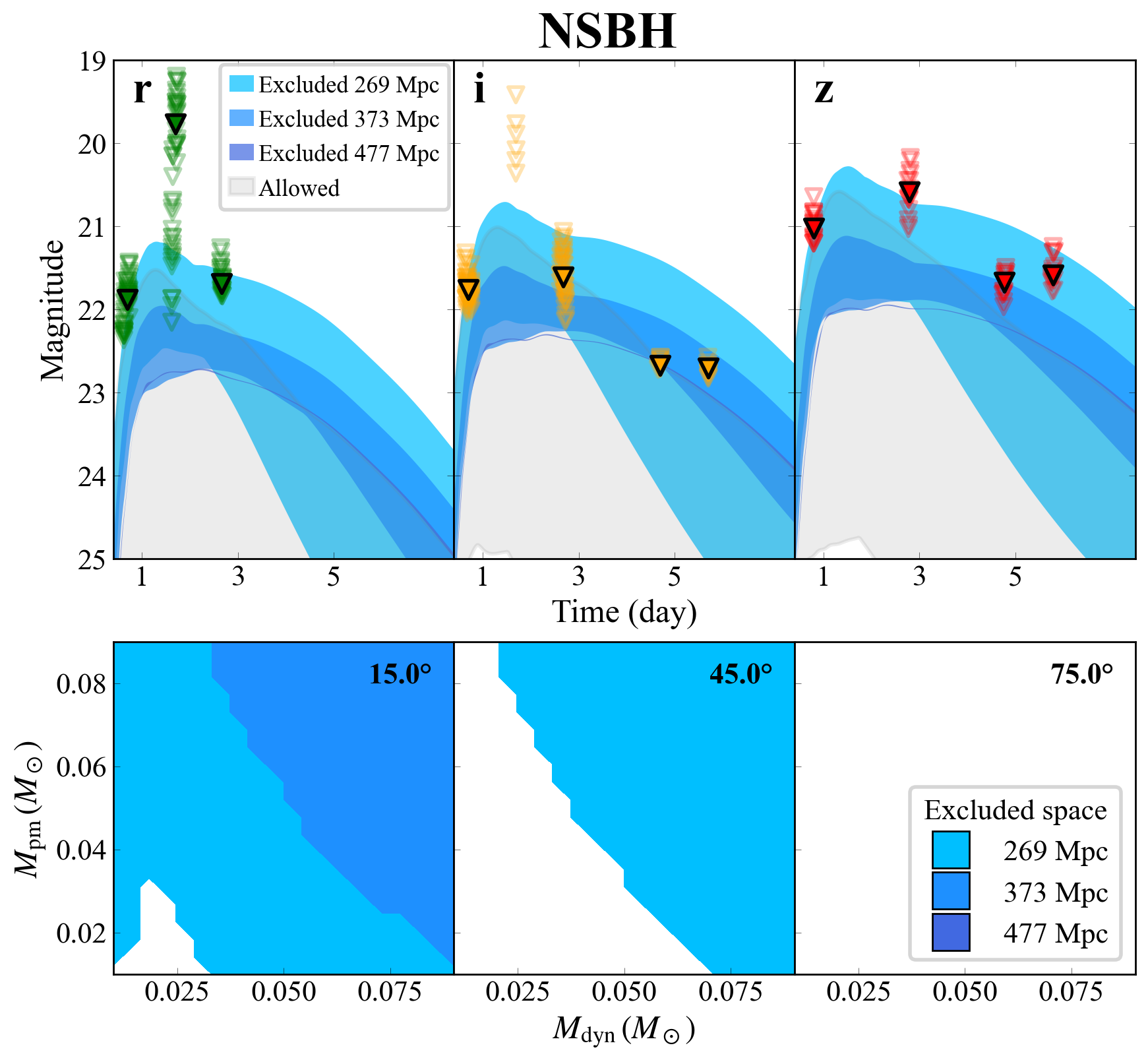}
        \end{overpic}

    \end{minipage}
    
    \caption{Parameters constraints based on the non-detection result, assuming a BNS (left) or a NSBH merger (right) for S250206dm. Two top panels show constraints on KN luminosity derived using the \texttt{POSSIS} model at different distances (corresponding to the median and $\pm1\sigma$ distances from Bilby skymap). The 5$\sigma$ limiting magnitude of each WFST pointing and the median magnitudes are plotted with open and solid inverted triangles, respectively. Galactic extinction with a median $E(B-V) = 0.1$ is applied in all panels. Light/deep blue and gray regions represent excluded and allowed KN scenarios, respectively. 
    For the BNS merger, black dashed and dash-dot lines represent AT 2017gfo-like KNe generated using \texttt{POSSIS} at 269 and 373 Mpc, respectively. Two bottom panels show constraints on ejecta mass. Three viewing angles of $15^\circ$, $45^\circ$ and $75^\circ$ are adopted. The fitted ejecta masses of AT 2017gfo from \cite{Dietrich_2020} are plotted as a red star for the BNS merger. }
    \label{fig:POSSIS}
\end{figure}

For a BNS merger origin, the constraint results based on \texttt{POSSIS} are shown in left two panels of Figure \ref{fig:POSSIS}. As shown in the top-left panel of Figure \ref{fig:POSSIS}, lightcurves of the AT 2017gfo-like KN and other simulated KNe from various parameter combinations are compared with WFST detection upper limits after considering Galactic extinction. The constraint mainly depends on observations of the first three nights due to the fast evolution for KNe from BNS mergers. At 269 Mpc, an AT 2017gfo-like KN can be ruled out by our observations in both $r$ and $i$ bands and most parameter combinations can also be excluded. The corresponding ejecta mass of the excluded lightcurves are shown in the bottom-left panel of Figure \ref{fig:POSSIS}. As the viewing angle ($\theta_\text{obs}$) or distance ($D_L$) increases, KNe become fainter, leading to a broader range of allowed ejecta masses. Based on the ejecta mass estimate of AT 2017gfo \citep{Dietrich_2020}, an AT 2017gfo-like KN is ruled out at 269 Mpc with viewing angle $\theta_\text{obs}<45^\circ$. At $\theta_\text{obs}=45^\circ$ and $D_L = 269$ Mpc, the masses of post-merger and dynamical ejecta are constrained to $M_\text{pm} \lesssim 0.03\,M_\odot$ and $M_\text{dyn} \lesssim 0.01\,M_\odot$, respectively. These mass upper limits are consistent with the ejecta produced in a scenario that a BNS merger directly collapses into a BH.

For a NSBH merger origin, similar results to BNS are also shown in right two panels of Figure \ref{fig:POSSIS}. KN lightcurves with bright peaks but rapid decay or faint peaks but slow decay are ruled out by the first and the fifth night observations, respectively. The WFST observations can effectively constrain KN lightcurves up to the median distance of 373 Mpc. As shown in the bottom-right panel of Figure \ref{fig:POSSIS}, at $\theta_\text{obs}=45^\circ$ and $D_\text{L}=269$ Mpc, a total ejecta mass greater than $\sim0.1\,M_\odot$ is ruled out. In the optimistic scenario ($\theta_\text{obs}=15^\circ, D_\text{L}=269$ Mpc), the masses of post-merger and dynamical ejecta can be further constrained to $M_\text{pm} \lesssim 0.03\,M_\odot$ and $M_\text{dyn} \lesssim 0.03\,M_\odot$, respectively. \par

\begin{figure}
    \centering
    \begin{overpic}[width=0.5\textwidth]{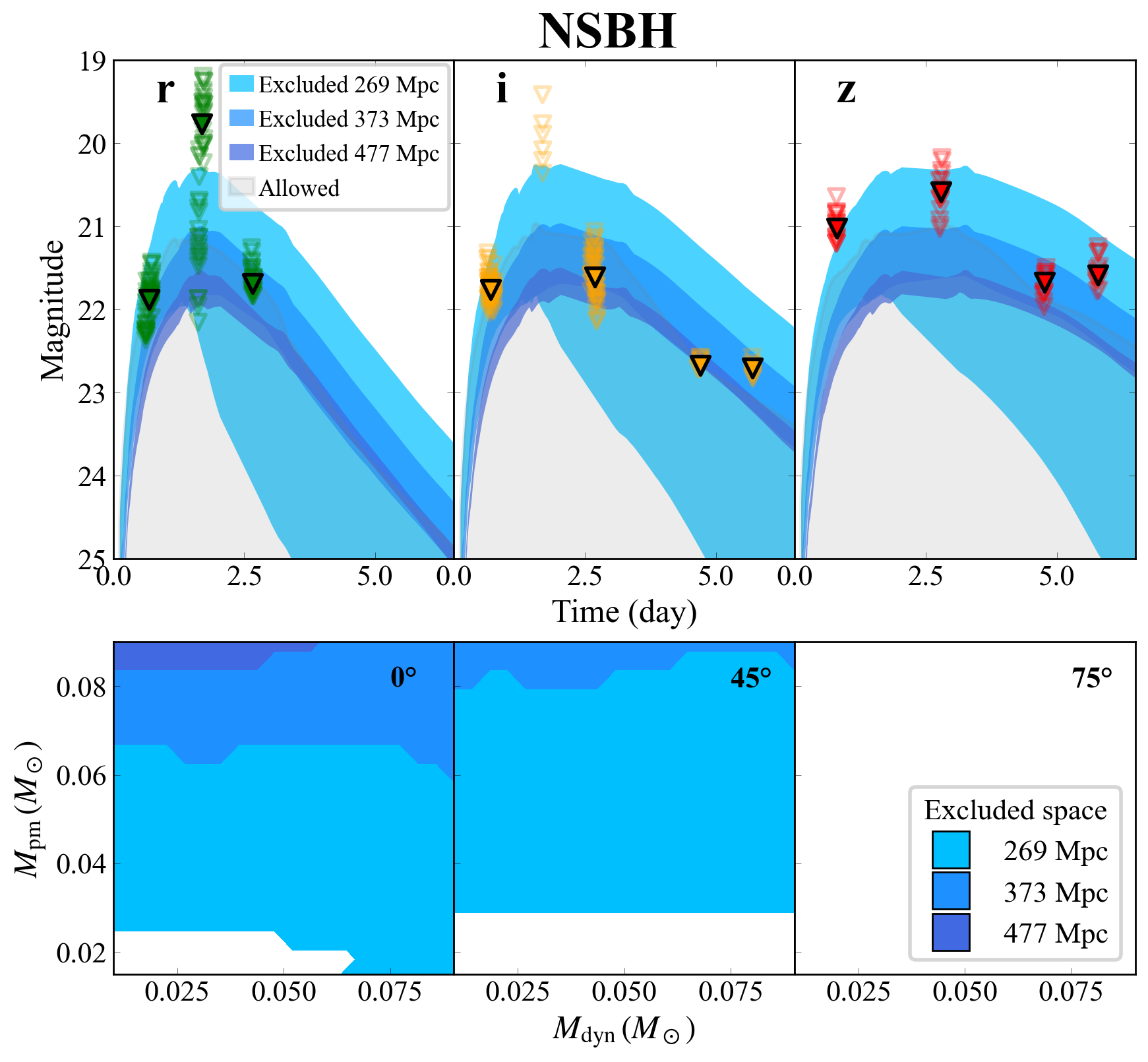}
    \end{overpic}
    \caption{Parameters constraint assuming a NSBH merger for S250206dm, similar to Figure \ref{fig:POSSIS}, but based on the \texttt{SuperNu} KN model. Top and bottom panels show the constraints on simulated KN lightcurves and ejecta masses produced during merger, respectively. The \texttt{SuperNu} NSBH model and three luminosity distances (corresponding to the median and $\pm1\sigma$ distances from Bilby skymap) are adopted. Galactic extinction with a median $E(B-V) = 0.1$ is adopted. In the top panel, the 5$\sigma$ limiting magnitude of each WFST pointing and the median magnitudes are plotted with hollow and solid inverted triangles, respectively. In the bottom panel, three viewing angles of $0^\circ$, $45^\circ$ and $75^\circ$ are adopted.}
    \label{fig:SuperNu}
\end{figure}

In comparison with \texttt{POSSIS}, we also employ another KN model \texttt{SuperNu} for the NSBH merger. \texttt{SuperNu} is a Monte Carlo radiative transfer code \citep{Wollaeger_2019,Wollaeger_2021,Korobkin_2021}, assuming a two-component axisymmetric ejecta that includes dynamical and wind ejecta (similar to poster-merger ejecta in \texttt{POSSIS}, hereafter also referred to as post-merger ejecta). A grid of simulation results is calculated using \texttt{SuperNu} with varying masses, velocities, morphologies, and compositions. In our calculations, a surrogate model for \texttt{SuperNu} from \cite{Kedia_2023} is adopted. Similar to the NSBH case in \texttt{POSSIS}, a morphology (named ``TSwind2'' in \cite{Kedia_2023}) is adopted for KNe produced by a NSBH merger, including a torus-shaped dynamical ejecta and a spherical post-merger ejecta. To reduce the model’s degrees of freedom, the velocities of the dynamical and post-merger ejecta are set to 0.18\,$c$ and 0.06\,$c$, respectively, corresponding to the average velocities derived from the velocity distribution of the \texttt{POSSIS} NSBH model. The ejecta mass ranges for the two components are both 0.01 to 0.09 $M_\odot$. The constraint results, similar to those in Figure \ref{fig:POSSIS}, are shown in Figure \ref{fig:SuperNu}. Compared with \texttt{POSSIS}, the trends of the results are similar, and the constraints weaken as the distance or viewing angle increases. For \texttt{SuperNu}, the KN lightcurves are more sensitive to $M_\text{pm}$: at a distance of 269 Mpc and a viewing angle of 45$^\circ$, $M_\text{pm} < 0.03\,M_\odot$ is derived from our observational upper limits, but with no effective constraint on $M_\text{dyn}$. Overall, the constraint on $M_\text{pm}$ is tighter than that for \texttt{POSSIS}, whereas the opposite holds for $M_\text{dyn}$. \par

\begin{figure}[htbp] 
    \centering
    \begin{overpic}[width=0.35\textwidth]{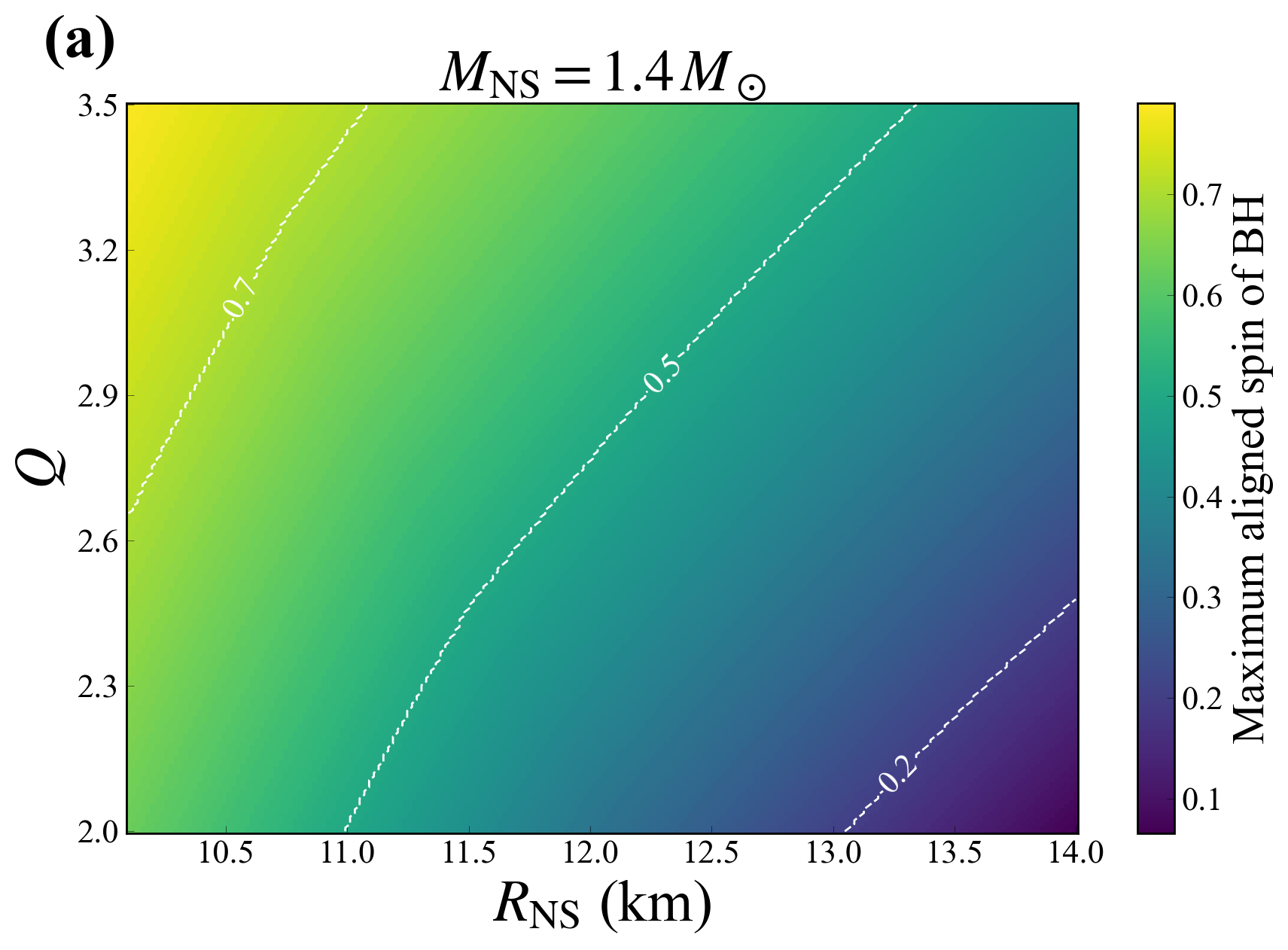}
    \end{overpic}
    \begin{overpic}[width=0.35\textwidth]{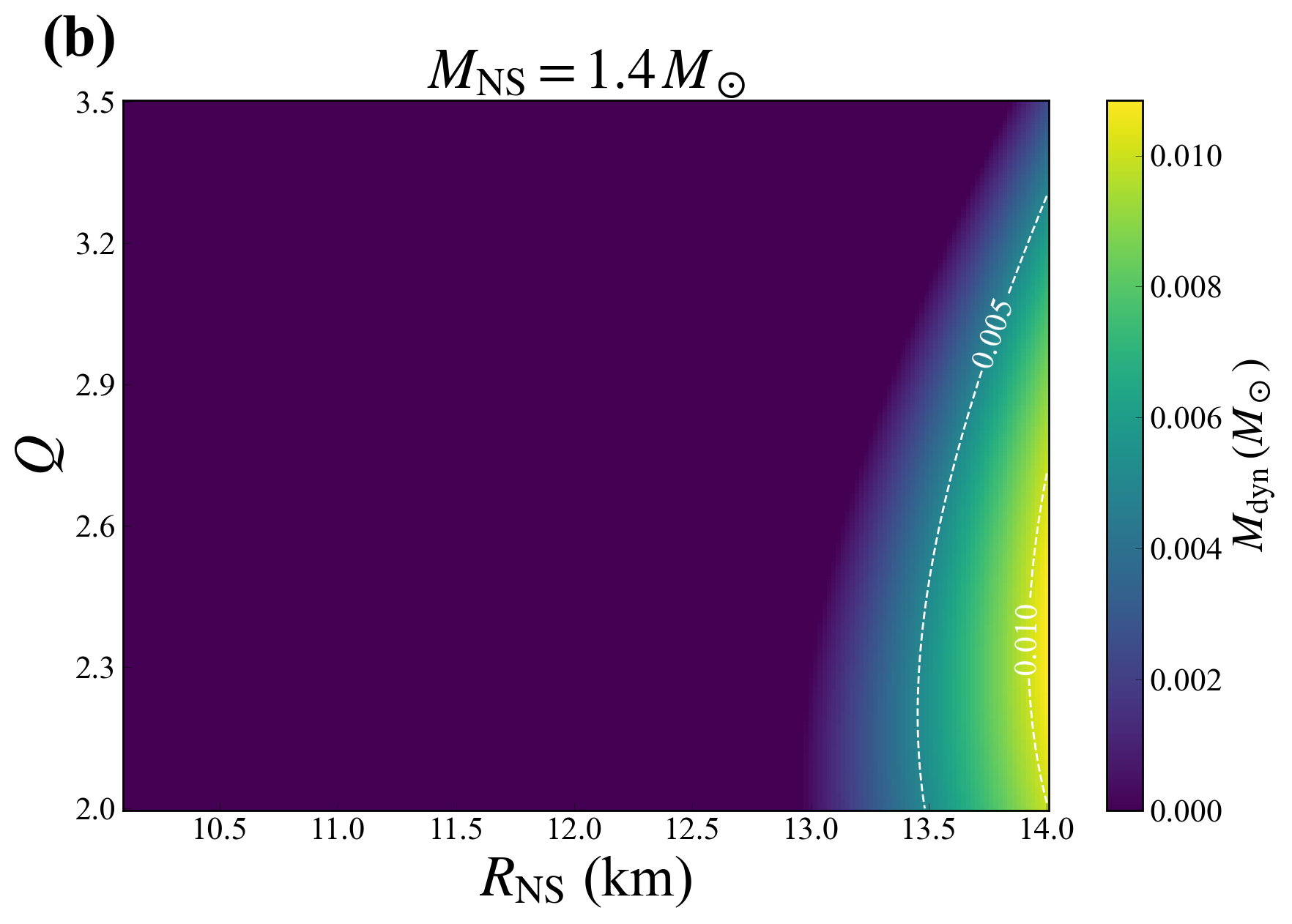}
    \end{overpic}
    \begin{overpic}[width=0.35\textwidth]{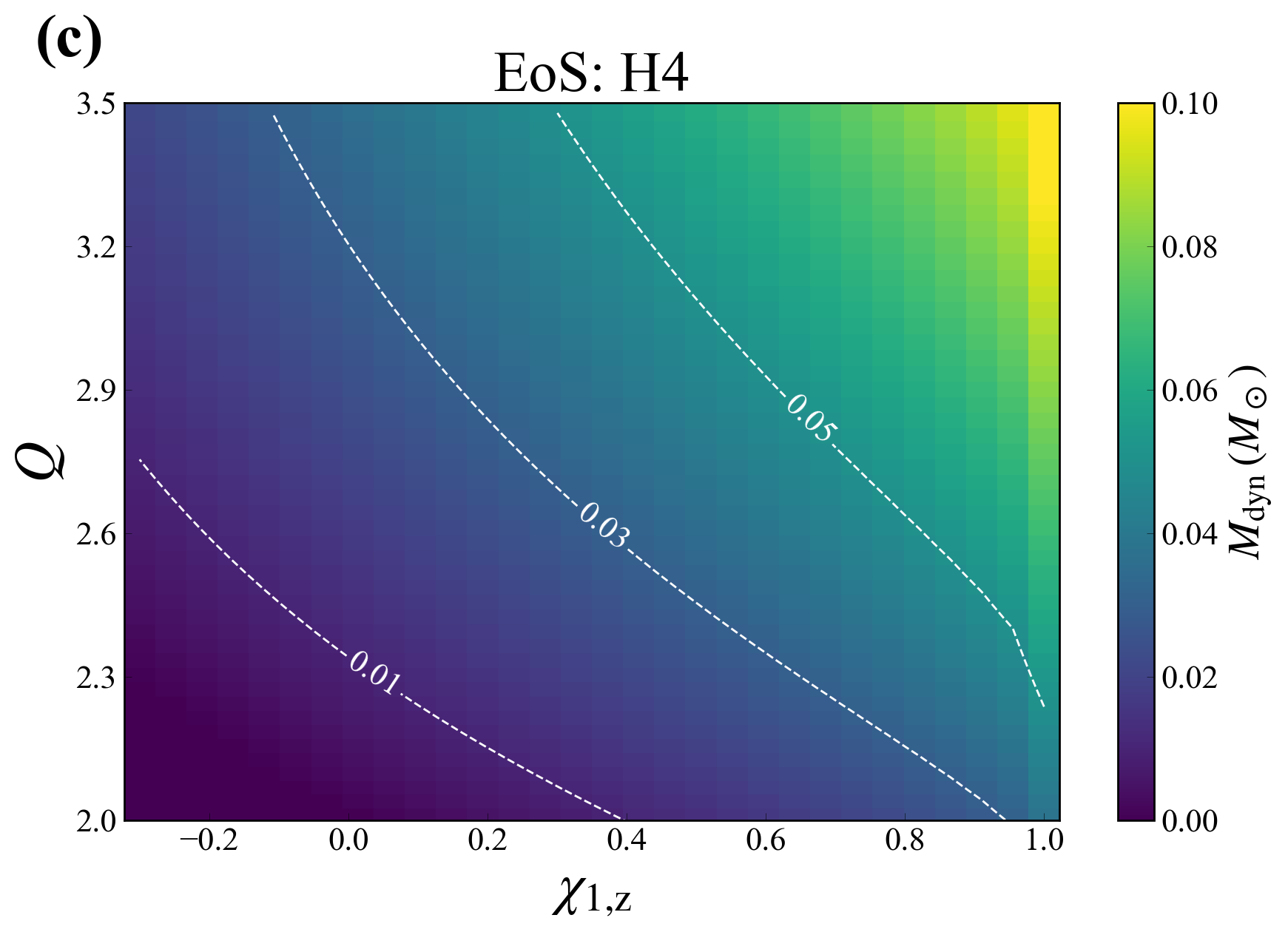}
    \end{overpic}
    \begin{overpic}[width=0.35\textwidth]{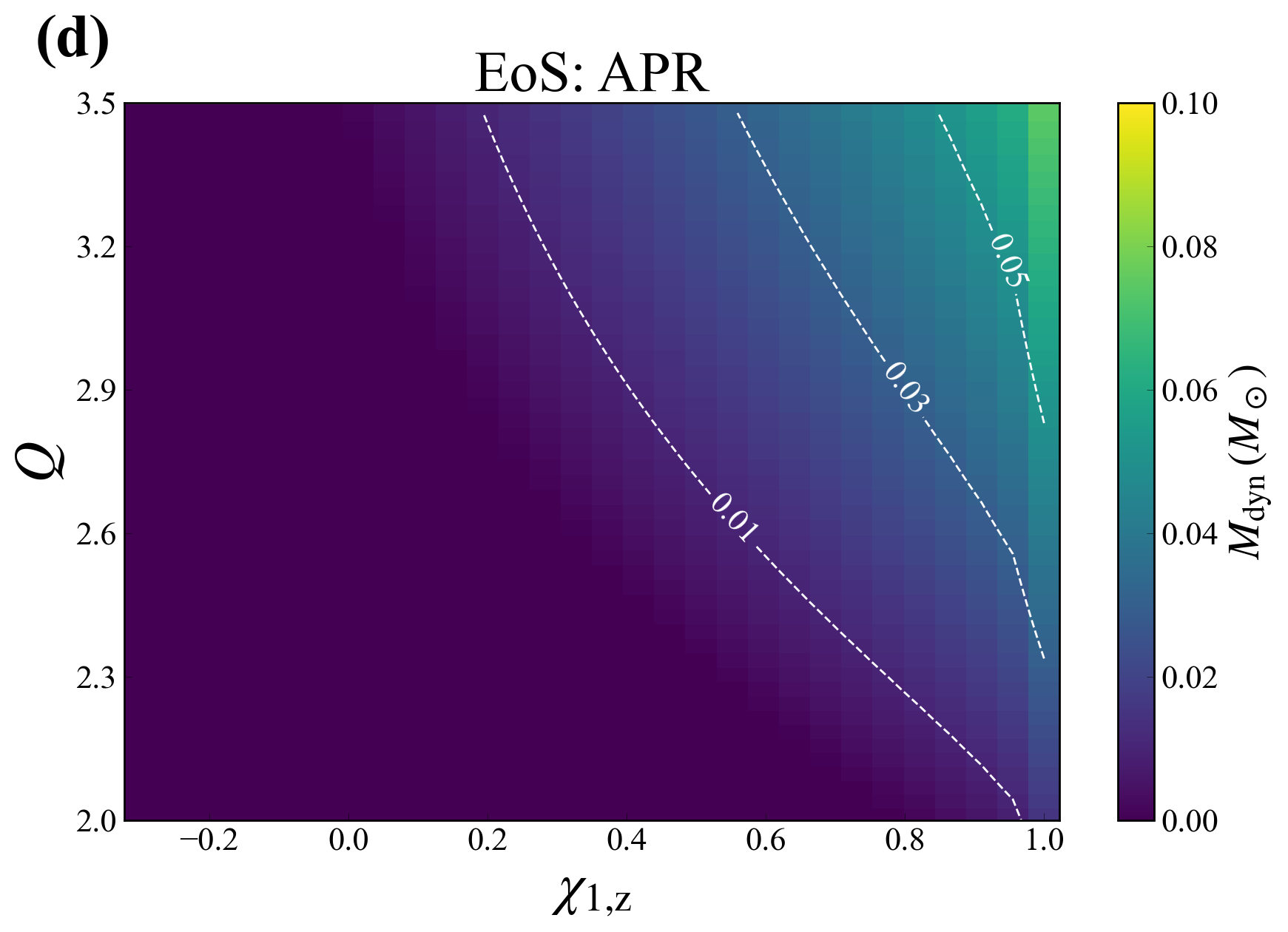}
    \end{overpic}
    \caption{The non-detection constraints on the mass ratio $(Q)$ and aligned spin of BH $(\chi_\text{1,z})$ for S250206dm. In Panel (a) and (b), the NS mass is set as $1.4\,M_\odot$, and the white dashed lines represent the contour of allowed maximum aligned spin of BH and the mass of dynamical ejecta, respectively. In Panel (a), the maximum aligned spin of BH derived from ejecta mass limits, where an optimistic result of $M_\text{wind} \lesssim 0.03\,M_\odot$ and $M_\text{dyn} \lesssim 0.03\,M_\odot$ is adopted. In Panel (b), a BH with zero aligned spin in system is adopted. In Panel (c) and (d), the EoSs of H4 and APR are adopted to calculate the produced dynamical ejecta mass, respectively. The white dashed lines represent the contour of the mass of dynamical ejecta produced during merger. The system chirp mass is set as $1.8\,M_\odot$, derived from source classification probability of S250206dm.}
    \label{fig:spin_q_NS_constrain} 
\end{figure}

\begin{figure}[htbp]
    \centering
    \begin{overpic}[width=0.5\textwidth]{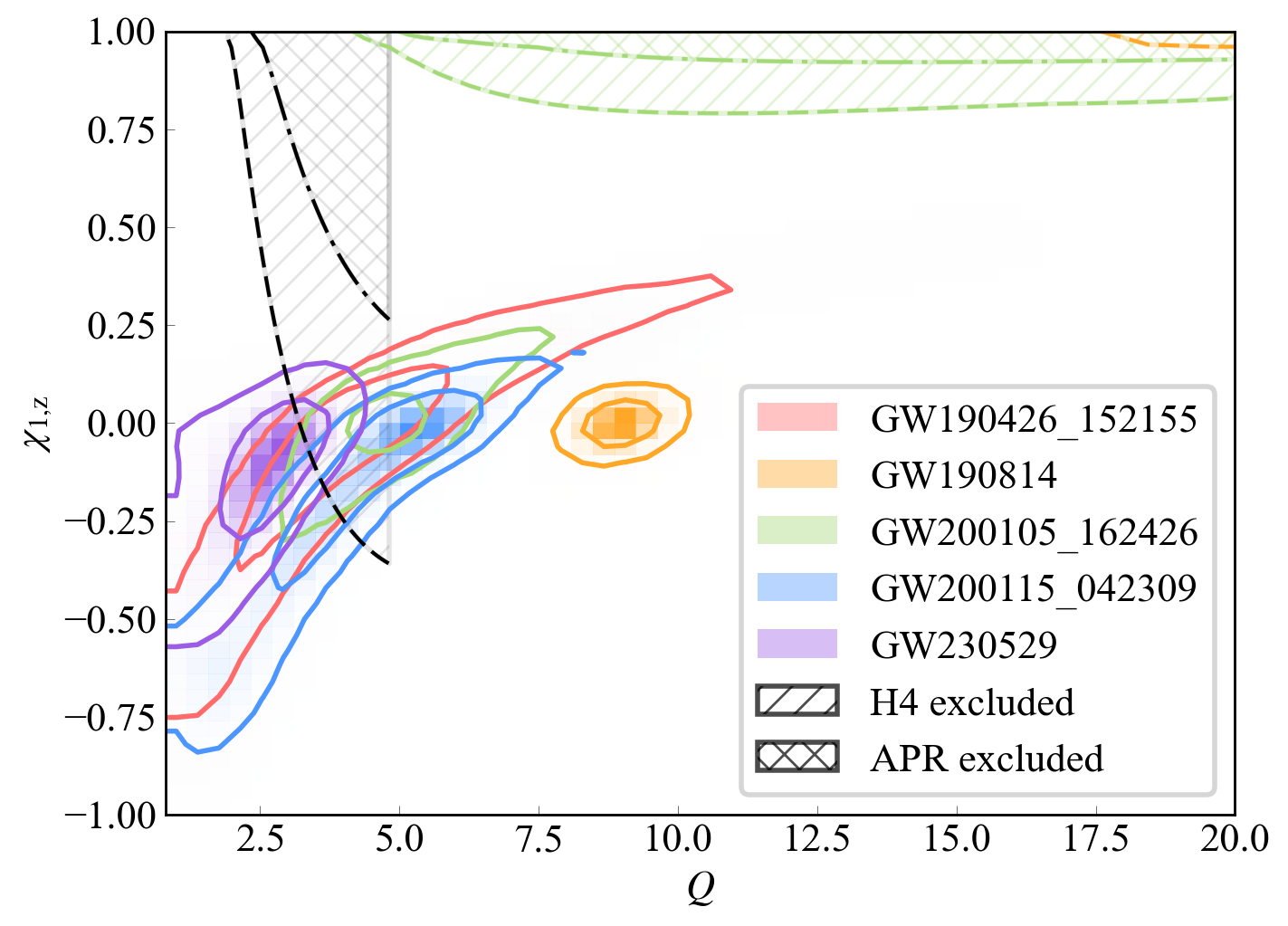}
    \end{overpic}
    \caption{The constraints on merger system from optical follow-ups compared with GW posterior estimates. With the chirp mass $(M_\text{c})$ fixed, the shaded areas with diagonal and cross represent the excluded spaces of the mass ratio $(Q)$ and aligned spin of BH $(\chi_\text{1,z})$ under EoSs of H4 and APR, respectively. $Q$ is restricted by the minimum NS mass assumed as $1\,M_\odot$. The gray shaded areas are ruled out space for S250206dm in this work, assuming $M_\text{c}$ as $1.8\,M_\odot$ and an optimistic constraint of $M_\text{dyn} \lesssim 0.03\,M_\odot$. For GW190814 and GW200105\_162426, the constrains are also shown based on the optimistic upper limits on $M_\text{dyn}$ \citep{Anand_2020,Andreoni_2020a} and $M_\text{c}$ estimates \citep{Abbott_2021,Abbott_2023} of ($\lesssim 0.001\,M_\odot$, $6.1\,M_\odot$) and ($\lesssim 0.02\,M_\odot$, $3.4\,M_\odot$), respectively. The contours represent the published GW posterior estimates of NSBH candidates in 50\% and 90\% confidence \citep{Abbott_2021,Abbott_2023,Abac_2024}.   
    }
    \label{fig:Chi-Q_Exclude}
\end{figure}

Combined with the numerical simulation results of BNS and NSBH mergers, the derived constraints on ejecta mass can be used to further infer the properties of the progenitor \citep{Andreoni_2020a,Coughlin_2020b,Anand_2020}. For a BNS merger origin for S250206dm, a large chirp mass leads to direct collapse into a BH during the merger. It is challenging to infer the merger system due to the theoretically small amount of ejecta. Therefore, we consider only the NSBH merger case here. The empirical equations from numerical simulation results, as used in \cite{Gompertz_2023} and \cite{Pillas_2025}, are adopted here to derive $M_\text{dyn}$ and $M_\text{pm}$ for the progenitor. Specifically, Equation 9 in \cite{Kruger_2020a} is used to calculate $M_\text{dyn}$. The total ejecta mass, $M_\text{rem}$, is estimated using Equation 4 in \cite{Foucart_2018}, where $M_\text{rem} = M_\text{disk} + M_\text{dyn}$. The post-merger (wind) ejecta mass is obtained by a fraction $\xi$ of disk material converted into wind ejecta, where $\xi$ is related to the binary mass ratio and can be estimated using Equation 12 in \cite{Raaijmakers_2021}. In our calculations, the H4 and APR EoSs are adopted, with their $R_{1.4}$ and $M_\text{TOV}$ values being (13.7 km, 2.03 $M_\odot$) and (11.3 km, 2.19 $M_\odot$), representing a ``stiff'' and a ``soft'' EoS, respectively. 

Derived from WFST observations, the constraints on the aligned spin of the BH $(\chi_\text{1,z})$ and the system mass ratio $(Q=M_\text{BH}/M_\text{NS})$ are shown in Figure \ref{fig:spin_q_NS_constrain}. In Panel (a) and (b), the NS masses are both assumed as $M_\text{NS} = 1.4\,M_\odot$. An optimistic result of $M_\text{pm} \lesssim 0.03\,M_\odot$ and $M_\text{dyn} \lesssim 0.03\,M_\odot$ is used to derive the upper limit on $\chi_\text{1,z}$, corresponding to $\theta_\text{obs}=15^\circ$ and $D_\text{L}=269$ Mpc. The produced dynamical ejecta mass, given the mass ratio and NS radius, is shown in Panel (b), where a BH with zero aligned spin is assumed, according to low aligned spin measurements of BH for most NSBH merger candidates detected to date (see Table \ref{tab:GW_ejecta}). Compared with the constraint of $M_\text{dyn} \lesssim 0.03\,M_\odot$ from the follow-up observations, it is challenging to derive an informative constraint on mass ratio and NS radius only by WFST observations. 

When combined with the GW signal, similar results are shown in Panel (c) and (d) of Figure \ref{fig:spin_q_NS_constrain}, where the system chirp mass is set to $M_\text{c} = 1.8\,M_\odot$, estimated from source classification probability of S250206dm using the method in \cite{Pillas_2025}. The method is described as follows: Estimate of source classifications in some GW online search pipelines, such as PyCBC Live, is based on the method in \cite{Villa-Ortega_2022}. Estimation primarily depends on the measurement of the chirp mass, assuming an astrophysical origin of the event, a correct distance estimate, and a point estimate of the source's detector frame chirp mass by template matching. The latest source classifications estimated by PyCBC Live are 37\% and 55\% for a BNS merger and a NSBH merger, respectively. Therefore, the chirp mass of S250206dm is derived using the inverse process of the method in \cite{Villa-Ortega_2022}, combined with the event distance estimate. Assuming minimum and maximum NS masses of 1 and 3 $M_\odot$, respectively, a source chirp mass of $\sim1.8\,M_\odot$ is estimated for S250206dm. For the H4 \citep{H4_1991} and APR \citep{APR_1998} EoSs, the region with $M_\text{dyn} \gtrsim 0.03\,M_\odot$ is excluded based on the observation-derived constraint, indicating that a NSBH with a large mass ratio and a high aligned spin is disfavored by our observations. Even for a BH with a low aligned spin ($\chi_\text{1,z} = 0$), a large mass ratio ($Q\gtrsim3.2$) is still disfavored assuming a ``stiff'' EoS like H4. \par

To assess the ability of optical follow-ups in constraining the properties of merger systems, we compare the GW posterior distributions of $\chi_\text{1,z}$ and $Q$ with the constraints derived from optical observations as shown in Figure \ref{fig:Chi-Q_Exclude}. For NSBH mergers involving a massive black hole ($>5\,M_\odot$), such as GW190814 and GW200105\_162426, optical data provide limited constraints. In contrast, for events located within the mass gap like S250206dm, WFST observing constraints reach into the GW posterior of most NSBH candidates. This optical-derived constraint achieves, for the first time, a precision comparable to that inferred from the GW signal, thereby complementing the GW estimate. The result shows that follow-up observations with limiting magnitudes of 22–23 mag can effectively constrain the mass ratio, aiding in distinguishing between a heavy BNS and a NSBH system for mass-gap events. \par 

\section{Other ToO observations and joint constraint} \label{sec_6}
\begin{figure}[htbp]
    \centering
    \begin{overpic}[width=0.6\textwidth]{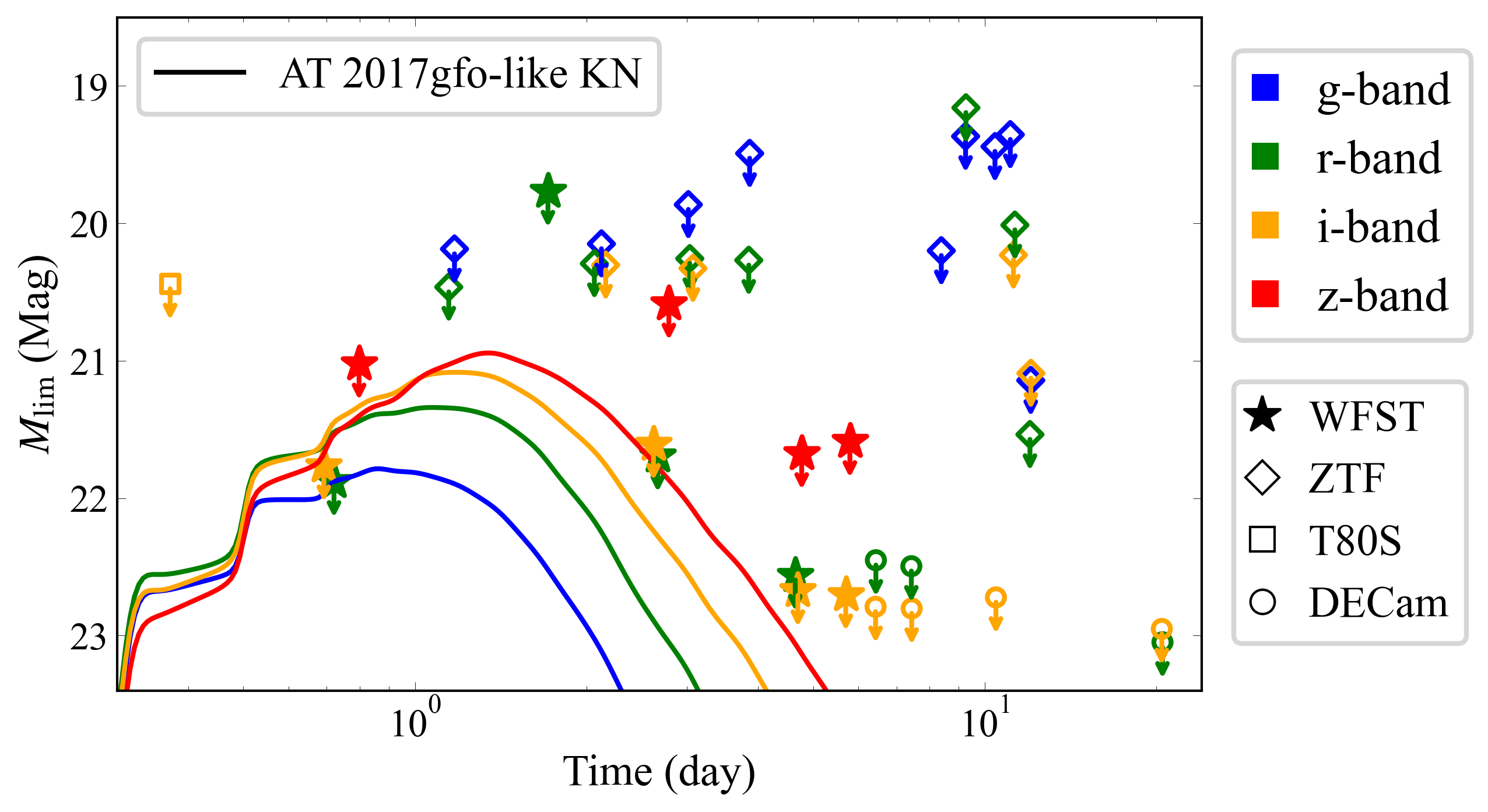}
    \end{overpic}
    \caption{Survey depths of WFST and other optical wide-field survey facilities for S250206dm. Lightcurves of AT 2017gfo-like KNe at 269 Mpc and median apparent 5$\sigma$ limiting magnitudes in different bands are represented by solid lines and scatter, respectively. The ZTF data are reported in \cite{Ahumada_2025}, and the T80S and DECam data are reported in \cite{Hu_2025}. Although the regions covered by DECam and T80S focus on the skymap in the southern sky and do not overlap with ZTF and WFST, their depths are also plotted for comparison. Galactic extinction is not considered here.}
    \label{fig:maglim_compare}
\end{figure}

Due to its importance and rarity, for S250206dm, ground-based wide-field survey facilities around the world participated in the EM counterpart search \citep{Hu_2025,Ahumada_2025,2025GCN.39211....1S,2025GCN.39241....1P,2025GCN.39256....1S}, such as ZTF and DECam. However, no valid EM counterpart was reported in the community. The information of some published follow-up observations is listed in Table \ref{tab:GW_ejecta}, including ZTF \citep{Ahumada_2025} and DECam \citep{Hu_2025}. WFST observations covered a comparable area (64\%) to ZTF (68\%), but with a detection limit (-16.1 mag) which is $\sim1.3$ magnitudes deeper than that of ZTF (-17.4 mag). Although WFST's limiting magnitude is shallower than that of DECam (-15.2 mag), its coverage is more timely and significantly wider compared to DECam, which began $\sim6$ days after the GW detection and covered a 10\% area. Additionally, at 269 Mpc, AT 2017gfo-like KNe can be excluded only by WFST observations, as shown in Figure \ref{fig:maglim_compare}. To derive the detection efficiency for AT 2017gfo-like KNe at different positions, we performed a quantitative analysis by randomly sampling $\sim$19,000 positions (RA, Dec and luminosity distance) according to the Bilby skymap \citep{2025GCN.39231....1L} of S250206dm using the \texttt{dynesty} sampler \citep{2004AIPC..735..395S,2020MNRAS.493.3132S,sergey_koposov_2025_17268284}. Detection is determined by comparing the luminosity of simulated KNe and 5$\sigma$ limiting magnitudes of WFST observations. Under the requirement of at least one detection in any band, the derived detection ratios under conditions without and with Galactic extinction are $\sim$34\% and $\sim$10\%, respectively. Therefore, in terms of both depth and coverage, WFST has been conducted the best follow-up observations of S250206dm to our knowledge, thanks to the powerful capability of WFST and the great observing conditions at Lenghu site \citep{Deng_2021}. \par

\begin{sidewaystable}
    \centering
    \begin{threeparttable}
    \small
    \renewcommand{\arraystretch}{1.2}
    \setlength{\tabcolsep}{1.5pt}
    \caption{\noindent\textbf{Properties of S250206dm and other published NSBH merger events or candidates.}}
    \begin{tabular}{l*{10}{c}}
        \toprule 
        Event\tnote{1} & $m_1$ & $m_2$ & $\chi_{1z}$ & $D_\text{L}$ & $M_{\rm dyn}$\tnote{2} & $M_{\rm wind}$ & $M_\text{lim,abs}$\tnote{3}& $P_\text{c}$ & Ref \\ 
        & $(M_\odot)$ & $(M_\odot)$ &  & (Mpc) & $(M_\odot)$ & $(M_\odot)$ & (mag) &  \\
        \midrule
        GW190426\_152155 & $6.2^{+1.8}_{-1.6}$ & $1.6^{+0.4}_{-0.3}$ & $-0.04^{+0.19}_{-0.29}$ & $370^{+180}_{-160}$ & -- & -- & -16.4 & 52\% &\cite{Kasliwal_2020}\\
        GW190814 & $24.4^{+0.7}_{-0.6}$ & $2.72^{+0.05}_{-0.05}$ & $-0.00^{+0.02}_{-0.02}$ & $240^{+40}_{-50}$ & -- & --& -14.9 & 99\% &\cite{Ackley_2020}\\
        GW191219\_163120 & $34.7^{+1.3}_{-1.4}$ & $1.30^{+0.04}_{-0.03}$ & $-0.00^{+0.04}_{-0.05}$ & $550^{+250}_{-160}$ &--&--&--&--&--\\
        GW200105\_162426 & $9.6^{+0.6}_{-0.7}$ & $2.02^{+0.11}_{-0.09}$ & $-0.00^{+0.05}_{-0.06}$ & $270^{+120}_{-110}$&--&--& -17.2 & 52\% &\cite{Anand_2020}\\
        GW200115\_042309 & $6.3^{+1.3}_{-1.8}$ & $1.53^{+0.48}_{-0.20}$ & $-0.15^{+0.16}_{-0.38}$ & $290^{+150}_{-100}$&--&--& -15.8& 22\% &\cite{Anand_2020}\\ 
        GW230529\_181500 & $3.8^{+0.51}_{-0.75}$ & $1.49^{+0.31}_{-0.15}$ & $-0.11^{+0.12}_{-0.19}$ & $201^{+102}_{-96}$ &--&0.001(--)&-16.0&16\% &\cite{Pillas_2025}\\
        \midrule
        \multirow{6}{*}{S250206dm\tnote{4}} & 3.84 & 1.20 & 0 & \multirow{6}{*}{$373^{+104}_{-104}$} & 0.029(--)&0.009(0.001)& \multirow{6}{*}{-16.1} & \multirow{6}{*}{64\%} & \multirow{6}{*}{WFST (this work)}\\ 
        & 3.19 & 1.40 & 0 &  & 0.008(--)  & 0.010(--)\\ 
        & 2.73 & 1.60 & 0 &  & --  & 0.004(--)\\  
        & 3.84 & 1.20 & 0.8 & & 0.070(0.038)&0.070(0.052)&\\
        & 3.19 & 1.40 & 0.8 & & 0.035(0.010)&0.087(0.059)&\\
        & 2.73 & 1.60 & 0.8 & & 0.008(--)&0.089(0.052)&\\
        \midrule
        \multirow{3}{*}{S250206dm} & & & & & & & -17.4 & 68\% & ZTF \citep{Ahumada_2025}\\
                                   & & & & & & & -15.2 & 9.3\% & DECam \citep{Hu_2025}   \\
                                   & & & & & & & -17.5 & 4.9\% & T80S \citep{Hu_2025} \\
        \bottomrule
    \end{tabular}
    \begin{tablenotes}
        \footnotesize
        \item[1] The information of published GW events are from the posterior distribution results in \cite{Abbott_2021}, \cite{Abbott_2023}, and \cite{Abac_2024}.
        \item[2] The ejecta mass produced are calculated assuming the EoS of H4(APR). $R_{1.4}$ and $M_\text{TOV}$ for EoSs H4 and APR are (13.7 km, 2.03 $M_\odot$) and (11.3 km, 2.19 $M_\odot$), respectively. Marker ``--" represents the produced ejecta less than $1\times10^{-3}\,M_\odot$.
        \item[3] The best median limiting magnitudes from multiple bands and the first three nights are adopted. The median limiting magnitude of DECam on the seventh night is adopted, as observations began on that night.
        \item[4] The chirp mass of $\sim 1.8M_\odot$ is adopted, estimated based on the source classification probability of S250206dm (see Materials and Methods). Assuming a non-rotating NS, $M_\text{TOV}$ is used to distinguish between NS and BH in our calculations.
    \end{tablenotes}
    \label{tab:GW_ejecta}
    \end{threeparttable}
\end{sidewaystable}

Combining the results of other ToO observations helps constrain the potential KN counterpart due to complementarity in observation times and covered region of different telescopes. For S250206dm, the ZTF observations effectively supplement the monitoring of the northern skymap one to two days after the merger, when WFST has relatively shallow depth due to cloudy weather (as shown in Figure \ref{fig:maglim_compare}). Combined with ZTF data, the joint constraints on the KN luminosity for BNS and NSBH mergers are shown in Panel (a) and (b) in Figure \ref{fig:joint_ZTF} in the appendix, respectively. For the BNS merger, some bright scenarios are excluded by the ZTF data. The strengthening of constraint is limited for both BNS and NSBH merges due to the difference in depth between WFST and ZTF. Additionally, the joint constraint combining with DECam and T80S data is inappropriate due to the lack of overlap in the covered region with WFST. However, these observations increase the reliability (covered probability increases from 64\% to 74\%) of the assumption in the model constraint that the merger occurred within the observed field.  \par

\section{Conclusion}  \label{sec_7}
In this work, we conducted timely follow-up observations by WFST and detailed EM counterpart search for the GW event S250206dm, which is the first well-localized neutron star merger candidate potentially located in the mass gap. Lasting for a week, WFST observations covered $\sim$64\% of the localization region in $r$, $i$, $z$ bands with a $5\sigma$ limiting magnitude up to 23 mag. After data reduction and candidate search, 12 off-nucleus extragalactic transients are discovered, but they are unlikely to be the KN or the afterglow related to S250206dm based on the analysis of host and photometric evolution. Compared with other ToO observations, the WFST follow-up for S250206dm is superior in both depth and coverage.

According to the observation limit, an AT 2017gfo-like KN counterpart to S250206dm at 267 Mpc can only be excluded by WFST. Combined with the KN model POSSIS, the ejecta mass is constrained as: $M_\text{pm} \lesssim 0.03\,M_\odot$ and $M_\text{dyn} \lesssim 0.01\,M_\odot$ for a BNS merger; $M_\text{pm} \lesssim 0.03\,M_\odot$ and $M_\text{dyn} \lesssim 0.03\,M_\odot$ for a NSBH merger, assuming the condition of a near on-axis observation and at 269 Mpc. Furthermore, the mass ratio of progenitor and aligned spin of BH are also constrained for NSBH merger. Compared with the estimate from GW signal, the constraint on the mass ratio through the optical follow-up can reach the similar order in accuracy for the first time. Our results show that rapid, deep follow-up observations are capable for constraining the properties of compact binary progenitors, providing crucial constraints on the nature of the mass gap.

The WFST observations of S250206dm indicate the potential capacity of WFST in searching for a KN counterpart \citep{Liu_2023}. Due to its unique geographical location, WFST can enhance the counterpart search of GW skymaps in the Northern Hemisphere. Combining with ZTF and Pan-STARRS, WFST is able to complement the wide-field survey facilities in the Southern Hemisphere, such as DECam and LSST. Looking forward, with the participation of WFST and LSST in GW EM counterpart search in upcoming O5\footnote{\href{https://observing.docs.ligo.org/plan/}{https://observing.docs.ligo.org/plan/}}, breakthroughs on compact binary merger within mass gap are expected to be achieved. \par

\begin{contribution}
Z.Y.L. conducted most of the analyses and wrote the manuscript under the supervision of W.Z.. W.Z., Z.P.J., and Z.G.D. initiated this study. Z.L.X., J.A.J., L.L.F., M.X.C., and L.H. contributed to the data reduction pipeline of Wide Field Survey Telescope (WFST). J.A.J., Z.Y.L., Z.L.X., W.Y.W., D.Z.M., J.H.Z., K.X.Y., and Z.Q.J. contributed to the human inspection of the candidate search. R.D.L., L.H., J.J.G., and D.X. helped and discussed the model analyses. T.G.W., X.K., X.F.W., Q.F.Z., Y.Q.X., J.A.J., N.J., Y.X.Z., H.F.Z., M.L., D.Z.Y., B.L., X.Z.Z., F.L., H.L., X.L.Z., J.L.T., H.R.W., Z.W., and J.W. contributed to the development of WFST. All of the authors contributed to the discussion.
\end{contribution}

\section*{Acknowledgments}
The Wide Field Survey Telescope (WFST) is a joint facility of the University of Science and Technology of China, Purple Mountain Observatory. We appreciate the members of the WFST operation and maintenance team for their support. 
WZ is supported
by the National Natural Science Foundation of China (Grant No. 12325301 and 12273035), Strategic Priority Research Program of the Chinese Academy of Science (Grant No. XDB0550300), the National
Key R\&D Program of China (Grant No. 2021YFC2203102 and 2022YFC2204602), and Cyrus Chun Ying Tang Foundations. 
The authors also gratefully acknowledge the support provided by National Key Research and Development Program of China (2023YFA1608100).
%

\software{Astropy \citep{astropy:2013, astropy:2018, astropy:2022}; SciPy \citep{2020SciPy-NMeth}; Numpy \citep{harris2020array}; MatPlotLib \citep{Hunter:2007}; SWarp \citep{SWarp}; dustmap \citep{Green2018}; ligo.skymap \citep{Singer_2016_a, Singer_2016_b}; HOTPANTS \citep{HOTPANTS}; afterglowpy \citep{Ryan_2020}; dynesty \citep{2004AIPC..735..395S,2020MNRAS.493.3132S,sergey_koposov_2025_17268284}}

\appendix
\clearpage
\begin{table}[htbp]
    \centering
    \begin{threeparttable}
    \renewcommand{\arraystretch}{1.0}
    \setlength{\tabcolsep}{2pt}
    \caption{\noindent\textbf{The cross-checked results of reported candidates based on WFST follow-up data.}} 
    \begin{tabular}{*{6}{c}}
        \toprule 
        TNS Name\tnote{1} & Ra & Dec & Coverage & Detection  & Team    \\
        \midrule
        AT 2025bcq  & 36.31265  & 50.17796  & Yes       & Yes       & ZTF           \\
        AT 2025bda  & 354.33854 & 22.97904  & No        & No        & ZTF           \\
        AT 2025bay  & 7.504874  & 37.16833  & Yes       & No        & ZTF           \\
        AT 2025brm  & 359.26723 & 29.48768  & Yes       & Yes       & ZTF           \\
        AT 2025ben  & 356.32103 & 28.02299  & Yes       & Yes       & ZTF           \\
        AT 2025bro  & 356.59715 & 28.65744  & Yes       & No        & ZTF           \\
        AT 2025brn  & 2.03711   & 32.84554  & Yes       & No        & ZTF           \\
        AT 2025bcx  & 7.35227   & 38.69038  & Yes       & No        & ZTF           \\
        AT 2025bcw  & 21.33198  & 45.27197  & Yes       & No        & ZTF           \\
        AT 2025brp  & 358.69192 & 32.65047  & Yes       & No        & ZTF           \\
        AT 2025brl  & 1.90744   & 27.96404  & Yes       & Yes       & ZTF           \\
        AT 2025bcr  & 20.49131  & 48.97043  & No        & No        & ZTF           \\
        AT 2025bew  & 31.57170  & 53.01039  & Yes       & No        & Pan-STARRS    \\
        AT 2025bey  & 25.75487  & 45.46341  & No        & No        & Pan-STARRS    \\
        AT 2025bex  & 24.46251  & 45.33960  & No        & No        & Pan-STARRS    \\
        AT 2025bev  & 30.24625  & 52.32467  & Yes       & No        & Pan-STARRS    \\
        AT 2025bcc  & 32.02902  & 55.34248  & Yes       & No        & Pan-STARRS    \\
        AT 2025bbt  & 45.46907  & 51.12973  & No        & No        & Pan-STARRS    \\
        AT 2025bbo  & 24.32197  & 45.72550  & Yes       & No        & Pan-STARRS    \\
        AT 2025bbn  & 37.78844  & 50.73533  & Yes       & No        & Pan-STARRS    \\
        AT 2025bbm  & 36.96265  & 49.47591  & Yes       & No        & Pan-STARRS    \\
        AT 2025bwl  & 357.78800 & 34.19678  & No        & No        & Pan-STARRS    \\
        AT 2025bce  & 31.23239  & 54.79534  & Yes       & No        & Pan-STARRS    \\
        AT 2025bbl  & 7.12748   & 33.19370  & No        & No        & Pan-STARRS    \\
        AT 2025bmq  & 38.51758  & 54.57247  & Yes       & Yes       & WL-GW         \\
        AT 2025azn  & 39.82257  & 49.57253  & Yes       & No        & SAGUARO       \\
        AT 2025azm  & 2.03128   & 32.56575  & Yes       & No        & SAGUARO       \\
        \bottomrule
    \end{tabular}
    \begin{tablenotes}
        \footnotesize
        \item[1] The candidates with Dec $\leq 0^\circ$ are not included, which are far from the region covered by WFST.
    \end{tablenotes}
    \label{tab:other_candidates_check}
    \end{threeparttable}
\end{table}

\begin{figure}
    \centering
    \begin{overpic}[width=0.48\textwidth]{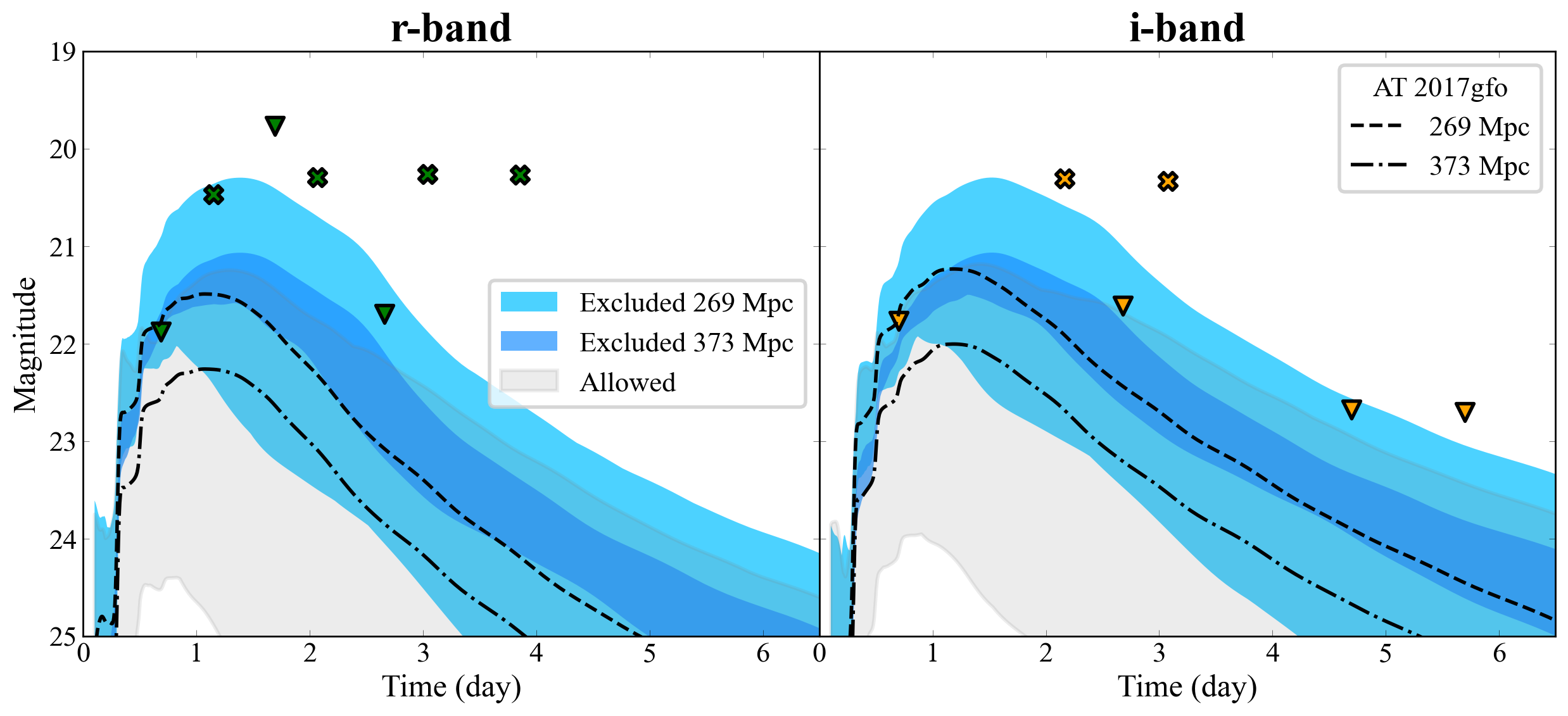}
    \put(0,45){{\textbf{(a)}}}
    \end{overpic}
    \begin{overpic}[width=0.48\textwidth]{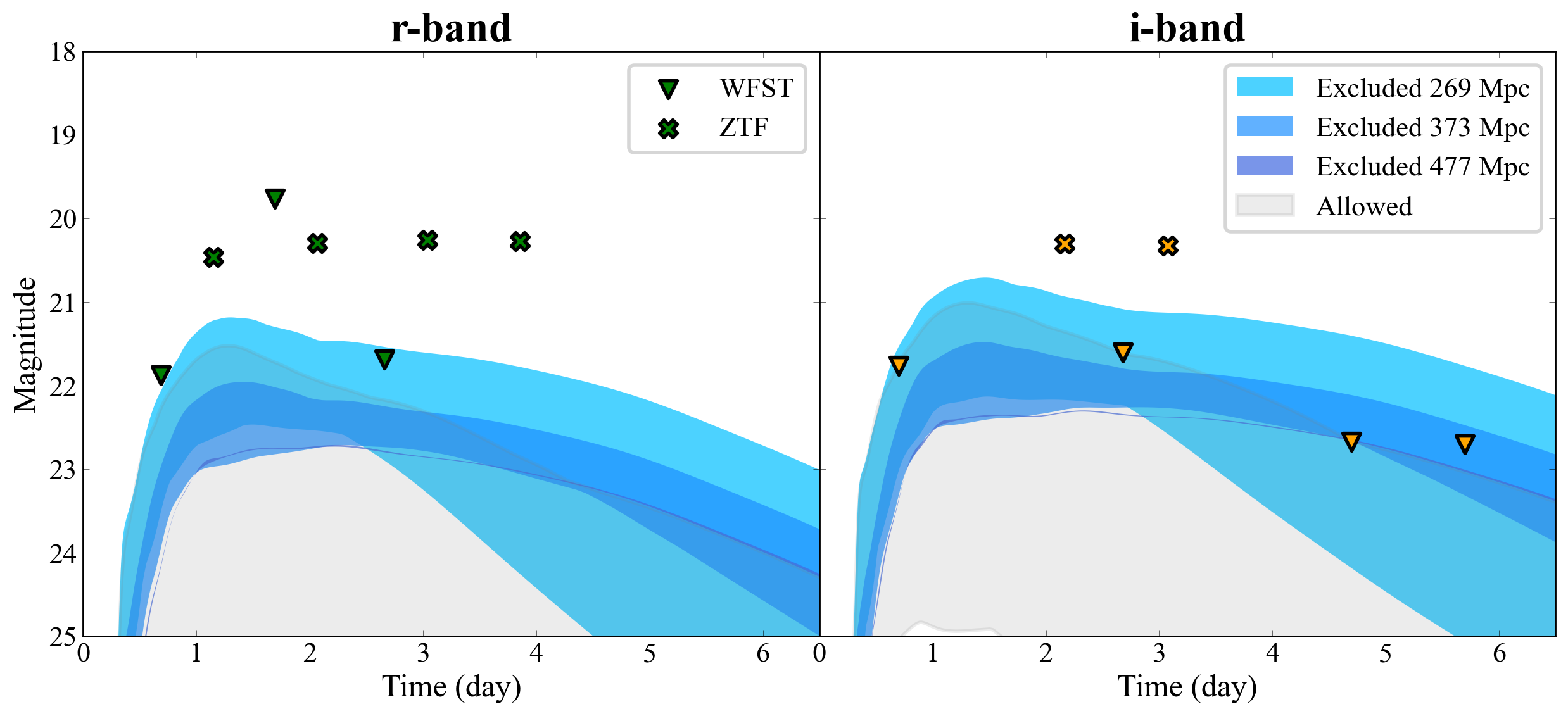}
    \put(0,45){{\textbf{(b)}}}
    \end{overpic}
    \caption{Joint constraints on KN luminosity, similar to top two panels in Figure \ref{fig:POSSIS}, but combining ZTF observations. Panel (a) and (b) correspond to BNS and NSBH mergers, respectively. The median limiting magnitudes of WFST and ZTF are represented by marks of inverted triangle and cross, respectively.}
    \label{fig:joint_ZTF} 
\end{figure}

\bibliography{draft}{}
\bibliographystyle{aasjournalv7}



\end{document}